\documentclass[,twocolumn,aps,prmaterials,amsmath,amssymb,longbibliography]{revtex4-2}

\usepackage{placeins}

\usepackage[T1]{fontenc}
\usepackage[utf8]{inputenc}
\usepackage{graphicx}
\usepackage{booktabs}
\usepackage{multirow}
\usepackage{siunitx}
\usepackage[caption=false]{subfig}
\usepackage{threeparttable}
\usepackage{xcolor}
\usepackage[expansion=false]{microtype}
\usepackage[colorlinks=true,linkcolor=blue!55!black,citecolor=blue!55!black,urlcolor=blue!55!black]{hyperref}
\usepackage{tikz}
\usepackage{newtxtext}
\usepackage{newtxmath}

\usepackage{orcidlink}

\usetikzlibrary{arrows.meta,positioning,shapes.geometric,fit,backgrounds,calc}

\graphicspath{{figures/}}

\newcommand{\Eform}{E_{\mathrm{f}}}
\newcommand{\Ehull}{E_{\mathrm{hull}}}
\newcommand{\Eg}{E_{g}}
\newcommand{\muBfu}{\ensuremath{\mu_{B}/\mathrm{f.u.}}}

\begin{document}

\title{Coordination-Resolved Surrogate Models for Thermodynamic Stability, Band Gaps, and Magnetic Moments of Spinel Oxides, Sulfides, and Selenides}

\author{Keltoum KHALLOUQ \orcidlink{0000-0002-5715-6351}}
\email{k.khallouq@gmail.com}
\affiliation{Laboratory of Solid Physics, Faculty of Sciences Dhar El Mahraz,
Sidi Mohamed Ben Abdellah University, Fez, Morocco}

\author{Ayoub EL MAAZOUZI \orcidlink{0000-0002-6181-5575}}
\email{ayb.elmaazouzi@gmail.com}
\affiliation{Laboratory of Solid Physics, Faculty of Sciences Dhar El Mahraz,
Sidi Mohamed Ben Abdellah University, Fez, Morocco}

\author{Rachid MASROUR \orcidlink{0000-0002-3646-665X}}  
\email{rachidmasrour@hotmail.com}
\affiliation{Laboratory of Solid Physics, Faculty of Sciences Dhar El Mahraz,
Sidi Mohamed Ben Abdellah University, Fez, Morocco}

\date{\today}
\date{July 04, 2026}

\begin{abstract}
The magnetism, electronic character, and thermodynamic stability of spinels
$AB_2X_4$ are set by how the cations distribute over the tetrahedral and
octahedral sublattices and by the local $d$-electron configuration on each.
We curated 320 cubic ($Fd\bar{3}m$) spinel entries from the Materials
Project-nitrides, oxides, sulfides, and selenides, including single-cation
mixed-valence $A_3X_4$ compounds-and trained tree-ensemble surrogates for
the formation energy, energy above the convex hull, band gap, total
magnetization, and metallicity. Cations were assigned to tetrahedral-like and
octahedral-like groups from CrystalNN coordination numbers rather than from
element identity, and the evaluation was group-aware throughout: splits were
grouped by reduced formula, every transform was fit on training folds only,
and champion models were selected on cross-validated scores before the
holdout was examined. Over twenty repeated grouped holdouts (single-seed
refits that reuse the tuned hyperparameters, and are therefore mildly
optimistic) the champions reach mean absolute errors of $0.121\pm0.030$
eV/atom for the formation energy, $0.048\pm0.013$ eV/atom for the hull
distance, and $1.27\pm0.19$~\muBfu{} for the magnetization, with a metallicity
accuracy of $0.85\pm0.06$. Band-gap regression does not beat a trivial
baseline on the 19-member non-metal holdout under paired bootstrap testing, we
report this negative result and trace it to sample scarcity and to the
semi-local DFT labels. On the identical grouped split, the tabular champion is
more accurate than an untuned MEGNet trained from scratch (0.087 versus 0.209
eV/atom formation-energy MAE on the primary holdout), a comparison that
bounds, rather than settles, the descriptor-versus-graph question at this
data scale. SHAP attribution ties the magnetization model to octahedral
$d$-occupancy and the formation-energy and band-gap models to
electronegativity descriptors, and grouped conformal intervals, permutation
nulls, and leave-one-chemistry-out tests bound a domain of applicability that
is uneven across anions and cations. The framework is a triage layer ahead of
targeted first-principles calculation, not a substitute for it.
\end{abstract}

\maketitle

\section{Introduction}

\subsection{Cation distribution in spinels}

The spinel structure, general formula $AB_2X_4$ and space group $Fd\bar{3}m$
(No.~227), places cations on two crystallographically distinct sites, the
tetrahedral $8a$ and the octahedral $16d$ positions, with anions at $32e$. The
partition of cations between the two sites is conventionally described by the
inversion parameter $\lambda$, which runs from the normal configuration
($\lambda=0$, all nominal $A$ cations tetrahedral) to the fully inverse one
($\lambda=1$)~\cite{PhysRev.98.391}. This occupancy matters physically: the edge-sharing $16d$
sublattice forms a pyrochlore-like network on which superexchange couplings,
whose sign and magnitude follow the Goodenough-Kanamori-Anderson rules~\cite{anderson1950, goodenough1955, kanamori1959}, compete with Zener double exchange in mixed-valence
compositions~\cite{zener1951}. Which mechanism dominates depends on which
cations, in which oxidation states, occupy which sites. Single-cation
mixed-valence spinels such as Fe$_3$O$_4$ and Co$_3$O$_4$ are the extreme
case: one element occupies both sublattices in different oxidation states, so
no site-assignment scheme based on element identity can describe them. The
thermodynamics of cation inversion, including its dependence on temperature
and synthesis route, has a long literature of its
own~\cite{oneill1983}.

Two points of scope should be stated at the outset. First, the dataset used
here consists of ordered DFT entries in space group 227, the tetragonal
$I4_1/amd$ branch is excluded, which removes Jahn-Teller-distorted ground
states such as hausmannite Mn$_3$O$_4$ (only cubic polymorphs of that
composition can enter the set). Ordered representations of inverse spinels are
often tabulated in lower-symmetry settings and may likewise be
underrepresented under this filter. Second, we do not compute $\lambda$;
the pipeline carries a binary occupancy proxy (Sec.~\ref{sec:feat}). The
featurization is therefore coordination-resolved, not inversion-resolved, and
we make no claim to model the continuous inversion degree.

\subsection{Target properties and motivation}

Spinel chalcogenides and oxides appear in permanent magnets, magnetic
recording media, photocatalysis, and spintronics, and their behavior ranges
from Mott and charge-transfer insulators to half-metals and ordinary band
metals depending on composition and site
chemistry. Previous computational studies have employed Monte Carlo simulations \cite{Khallouq2023, Maazouzi2019}, often parameterized from first-principles calculations, to investigate finite-temperature magnetic ordering and thermodynamic properties of selected spinel compounds. Experimental magnetization data remain
sparse compared with the size of the accessible $A$-$B$-$X$ space, which is
the practical argument for computational
pre-screening~\cite{jain2016review}. Exhaustive DFT relaxation of that space is
not feasible, so a cheap statistical surrogate that ranks candidates before
first-principles work has clear value, provided its errors and domain of
validity are quantified rather than assumed.

\subsection{This work}

We treat machine learning strictly as an interpolative surrogate on
physics-informed descriptors. The specific contributions are:
a coordination-based site assignment (CrystalNN) that processes normal,
single-cation, and mixed-occupancy entries in one pipeline, a group-aware
evaluation protocol, grouped by reduced formula, in which every transform is
fit on training folds only and champion models are chosen on cross-validated
scores before the holdout is consulted, conformal prediction intervals,
label-permutation nulls, and leave-one-anion/cation-out tests that bound the
domain of applicability, an explicit accounting of where the surrogate fails,
including a band-gap negative result, and a screening exercise with DFT-ready
export, presented as prioritization for first-principles validation. A
predictive claim is made below only where the model beats trivial and linear
baselines under paired bootstrap testing and survives a label-permutation
null.

\subsection{Relation to prior work}

Composition-only tabular surrogates built on Magpie-style
statistics~\cite{ward2016magpie} cover broad chemistry but cannot distinguish a
normal from an inverse spinel of identical composition, which is exactly the
degree of freedom that matters here. Graph neural networks pretrained on
repository-scale corpora (CGCNN, MEGNet, ALIGNN)~\cite{xie2018cgcnn,chen2019megnet}
are accurate on average but pose a benchmarking hazard for narrow
sub-families: when the evaluation compounds were part of the pretraining
corpus, the reported errors are contaminated upper bounds rather than
generalization estimates. Benchmarking studies have also documented that tree
ensembles on descriptors tend to beat graph networks below roughly $10^4$
training samples~\cite{dunn2020matbench}, and that accurate formation energies do
not automatically translate into accurate hull stabilities~\cite{bartel2020}.
Our contribution is not a new learner. It is a leakage-audited evaluation
protocol plus a site-resolved featurization for one structure family, with the
pretrained-GNN reference explicitly labeled as contaminated and a from-scratch
GNN trained on the identical split reported as the like-for-like comparison
(Sec.~\ref{sec:gnn}).

\section{Methods}

\subsection{Dataset curation and group-aware partitioning}
\label{sec:data}

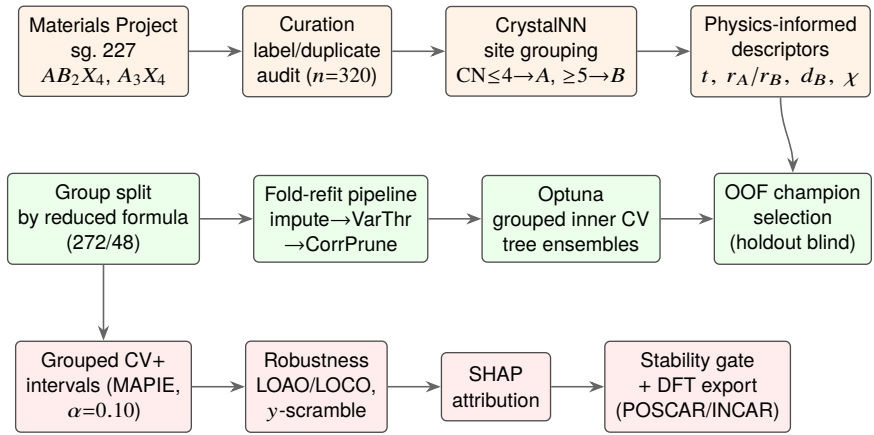
\begin{figure*}[htbp]
\centering
\footnotesize\sffamily
\begin{tikzpicture}[
  node distance=6mm and 7mm,
  box/.style={rounded corners=2pt,draw=black!70,line width=0.5pt,align=center,
              inner sep=4pt,minimum height=9mm,fill=blue!5,font=\footnotesize},
  data/.style={box,fill=orange!10},
  model/.style={box,fill=green!8},
  outb/.style={box,fill=red!7},
  ar/.style={-{Stealth[length=2mm]},draw=black!60,line width=0.6pt}]
  \node[data] (mp) {Materials Project\\sg.~227\\$AB_2X_4$, $A_3X_4$};
  \node[data,right=of mp] (curate) {Curation\\label/duplicate\\audit ($n{=}320$)};
  \node[data,right=of curate] (cn) {CrystalNN\\site grouping\\$\mathrm{CN}{\le}4{\to}A$, ${\ge}5{\to}B$};
  \node[data,right=of cn] (feat) {Physics-informed\\descriptors\\$t,\;r_A/r_B,\;d_B,\;\chi$};
  \node[model,below=10mm of mp] (split) {Group split\\by reduced formula\\(272/48)};
  \node[model,right=of split] (prep) {Fold-refit pipeline\\impute$\to$VarThr\\$\to$CorrPrune};
  \node[model,right=of prep] (tune) {Optuna\\grouped inner CV\\tree ensembles};
  \node[model,right=of tune] (sel) {OOF champion\\selection\\(holdout blind)};
  \node[outb,below=10mm of split] (unc) {Grouped CV$+$\\intervals (MAPIE,\\$\alpha{=}0.10$)};
  \node[outb,right=of unc] (rob) {Robustness\\LOAO/LOCO,\\$y$-scramble};
  \node[outb,right=of rob] (shap) {SHAP\\attribution};
  \node[outb,right=of shap] (screen) {Stability gate\\$+$ DFT export\\(POSCAR/INCAR)};
  \draw[ar] (mp)--(curate); \draw[ar] (curate)--(cn); \draw[ar] (cn)--(feat);
  \draw[ar] (feat.south) to[out=-90,in=90] (sel.north);
  \draw[ar] (split)--(prep); \draw[ar] (prep)--(tune); \draw[ar] (tune)--(sel);
  \draw[ar] (split.south) to[out=-90,in=90] (unc.north);
  \draw[ar] (unc)--(rob); \draw[ar] (rob)--(shap); \draw[ar] (shap)--(screen);
\end{tikzpicture}
\caption{Workflow. Curated Materials Project spinels are featurized through
CrystalNN site grouping and physics-informed descriptors (top row), then
partitioned and modeled under a group-aware protocol in which all transforms
are refit per fold and champions are selected on out-of-fold scores before the
holdout is seen (middle row). Conformal intervals, robustness audits,
attribution, and a gated stability screen with DFT-ready export follow (bottom
row).}
\label{fig:workflow}
\end{figure*}

Figure~\ref{fig:workflow} summarizes the full pipeline, from retrieval to the
gated screening export. Entries were retrieved from the Materials
Project~\cite{jain2013mp} for space group 227 through the \texttt{mp-api} summary
endpoint, the tetragonal branch (space group 141) was excluded in this run.
Entries lacking a reported formation energy would have been removed first, 
none occurred in this retrieval (rejection log archived with the result bundle). 
Areduced-formula stoichiometry filter then admitted the canonical $AB_2X_4$
composition and the single-cation $A_3X_4$ composition, rejecting 1059
entries, a further 62 entries failed the coordination-descriptor computation
of Sec.~\ref{sec:feat}. The curated set contains 320 entries. No explicit
anion filter was applied, the anions observed in the curated set are N, O, S,
and Se, with oxides dominating, smaller sulfide and selenide populations, and
a small nitride minority. The cation set is transition-metal-rich and also
contains lanthanide-bearing chalcogenides. All rejection counts and
identifiers are archived with the result bundle.

Labels are Materials Project quantities: formation energy per atom, energy
above the convex hull, band gap, and total magnetization. The magnetization
was converted to per formula unit by dividing the reported cell value by the
number of formula units (\texttt{nsites}/7). The binary metallicity label was
defined by thresholding the reported gap at \SI{0.01}{eV} wherever a gap is
reported (a sensitivity study over the threshold appears in
Sec.~\ref{sec:perf}), entries with a missing label were excluded from the
corresponding task and never imputed. Materials Project energies come from a
mixed GGA/GGA$+U$ scheme with compatibility
corrections~\cite{wang2021corrections}, "PBE" is shorthand below for that
scheme, and the gaps are semi-local Kohn-Sham gaps with the usual
underestimation~\cite{perdew1985}. A duplicate audit by reduced formula
returned zero duplicated compositions.

Because several entries can share one reduced formula (polymorphs,
re-relaxations), plain $K$-fold cross-validation would place near-identical
compositions in both training and evaluation folds. Partitioning therefore
used \texttt{GroupShuffleSplit} grouped by reduced formula, holdout fraction
0.15, seed 42, giving 272 training and 48 held-out entries with zero shared
formulas (verified by an intersection assert at run time). Table~\ref{tab:dataset}
summarizes the provenance.

\subsection{Local-environment featurization}
\label{sec:feat}

Cations were assigned to sublattice groups from CrystalNN~\cite{zimmermann2020}
coordination numbers: sites with $\mathrm{CN}\le 4$ enter the tetrahedral-like
$A$ group and sites with $\mathrm{CN}\ge 5$ the octahedral-like $B$ group.
Cation sites for which CrystalNN raises an exception are excluded from the
assignment, if either group ends up empty, the cations are instead split at
the median coordination number, and entries for which no valid split exists
are rejected (part of the 62 descriptor rejections above). Oxidation states
were assigned with pymatgen's guessing routines, with element-default states
as fallback.

On this assignment we computed: Shannon ionic radii $r_A^{IV}$, $r_B^{VI}$,
and $r_X^{VI}$ (when the radius for the preferred coordination is unavailable
the code falls back to the CN-VI value and then to the element's generic ionic
radius, so a small fraction of radii are not strict Shannon values at the
nominal coordination), the ratio $r_A/r_B$, the size-matching factor
\begin{equation}
t = \frac{r_A + r_X}{\sqrt{2}\,(r_B + r_X)};
\label{eq:tol}
\end{equation}
site-resolved mean oxidation states, formal $d$-electron counts $d_A$ and
$d_B$ (group number minus oxidation state, clamped to $[0,10]$) and their sum;
a transition-metal fraction, site- and anion-resolved electronegativities
$\chi_A$, $\chi_B$, $\chi_X$ and the anion-cation contrast
$\Delta\chi=\chi_X-\tfrac12(\chi_A+\chi_B)$, a Jahn-Teller flag that marks
octahedral cations with formal $d^4$ or $d^9$ counts (the spin state is not
resolved), the coordination-number gap $\mathrm{CN}_B-\mathrm{CN}_A$, and a
binary occupancy proxy set to one when the dominant $A$-group and $B$-group
elements coincide, as in single-cation spinels. These were combined with
Magpie compositional statistics (matminer \texttt{ElementProperty}, magpie
preset) and with CrystalNN site fingerprints (\texttt{SiteStatsFingerprint}
over the \texttt{CrystalNNFingerprint} "ops" preset, mean and standard
deviation per motif).

The raw feature matrix holds 264 descriptors: 120 Magpie statistics after
removal of the \texttt{SpaceGroupNumber} and \texttt{MendeleevNumber} families
(categorical proxies that could act as shortcuts), 19 bespoke structural and
electronic descriptors, a 3-descriptor occupancy block (Jahn-Teller flag,
occupancy proxy, coordination-number gap) used in the ablation of
Sec.~\ref{sec:shap}, and 122 CrystalNN fingerprint statistics. The energy
above the hull, density, volume, and the Materials Project stability flag are
outputs or near-outputs and were excluded from the feature matrix for every
task.

\subsection{Learning pipeline, tuning, and model selection}
\label{sec:pipe}

Every model is wrapped in a single scikit-learn pipeline~\cite{pedregosa2011sklearn}, median imputation
followed by a variance filter (threshold $10^{-10}$) and deterministic
correlation pruning (first-kept scan at $|r|>0.90$), so that all
transforms are refit on the training portion of every split, including inside
the tuning loop. Tree models receive the pruned features directly, only the
linear baselines standardize internally. The baselines are a
\texttt{DummyRegressor} predicting the training median (regression), a
\texttt{DummyClassifier} predicting the most frequent training class
(classification), \texttt{RidgeCV}, and logistic regression. Against these we
benchmarked XGBoost~\cite{chen2016xgboost}, LightGBM~\cite{ke2017lightgbm}, CatBoost~\cite{dorogush2018catboost}, and random forests.

Hyperparameters were tuned with Optuna (TPE sampler, 50 trials per target and
model) on a grouped three-fold inner cross-validation of the training
partition, minimizing MAE for regression and error rate for classification.
Champion selection used repeated grouped five-fold out-of-fold (OOF)
statistics on the training partition (four repeats with re-randomized fold
assignments), by mean OOF MAE for regression and mean OOF AUC for
classification, the holdout was not consulted at any stage of tuning or
selection. One caveat applies to the OOF numbers themselves: the tuning loop
ran on the same partition, and the OOF evaluation was not nested inside it, so
the OOF values reported in Table~\ref{tab:oof} are selection scores, mildly
optimistic as performance estimates. The clean generalization estimates in
this paper are the holdout and the repeated-holdout results.

On the primary holdout, every model (baselines included) was evaluated as a
three-member seed ensemble (seeds 42, 7, 123, predictions averaged). The
repeated-holdout, leave-one-chemistry-out, and ablation analyses of
Sec.~\ref{sec:robust} use single-seed fits, this asymmetry is quantified in
Sec.~\ref{sec:perf}.

For each regression champion, prediction intervals were built with MAPIE using
the "plus" method over a grouped five-fold split (grouped CV$+$), at
miscoverage $\alpha=0.10$. The construction assumes exchangeability between
calibration and test data, it does not hold in the extrapolation regime of
Sec.~\ref{sec:screening}, and we flag that explicitly there. The empirical
calibration of these intervals is examined in Sec.~\ref{sec:perf}.

\subsection{Statistical evaluation and robustness protocol}
\label{sec:robust}

Champion-versus-baseline MAE differences on the holdout were tested by paired
bootstrap (2000 resamples of the 48 holdout entries, 19 for the gap), with a
95\% percentile interval, a difference is called significant when the interval
excludes zero. Twenty additional grouped holdouts (15\% test fraction, seeds
100-119) were drawn over the full 320-entry set and evaluated with
single-seed fits of the dummy, ridge, and champion models. Two caveats apply:
these repeated splits reuse the hyperparameters tuned once on the primary
training partition, so most repeated test folds contain compounds present
during tuning, and the same reuse applies to the leave-one-anion-out (LOAO)
and leave-one-cation-out (LOCO) tests below. Both analyses are therefore
mildly optimistic and are read as robustness checks, not as fresh unbiased
estimates.

LOAO retrains the champion with one anion class withheld (evaluated when at
least 5 test and 20 training entries remain), LOCO does the same for a probe
set of nine cations (Al, Co, Cr, Cu, Fe, Mg, Mn, Ni, Zn) that appear as a
dominant $A$- or $B$-group element in at least ten entries, the full cation
content of the dataset is broader. A label-permutation null (five permutations
per regression target, grouped OOF protocol) checks that the pipeline cannot
manufacture skill from scrambled labels. Feature-group ablations
(Sec.~\ref{sec:shap}) use the grouped-OOF protocol with the champion
hyperparameters.

\subsection{Computational environment}
\label{sec:env}

The pipeline ran under Python 3.12 on Google Colab, the graph-network
baselines and the GPU-enabled boosters (XGBoost, CatBoost) used a single
NVIDIA T4. Pinned package versions: NumPy 1.26.4, scikit-learn 1.5.2, LightGBM
4.5.0, CatBoost 1.2.7, SHAP 0.46.0, MAPIE 0.9.1, gplearn 0.4.2, PyTorch 2.4.0,
DGL 2.4.0, MatGL 1.1.1. XGBoost, Optuna, pymatgen, matminer, and
\texttt{mp-api} were installed unpinned, the exact versions of every package
in the archived run are recorded in \texttt{run\_metadata.json} within the
result bundle, together with the Materials Project retrieval date. Fixed seeds
were used throughout (global seed 42, holdout ensemble seeds 42, 7, 123),
though GPU tree-boosting backends are not bit-reproducible across hardware.

\subsection{Graph-network baselines}
\label{sec:gnn}

Two kinds of graph-network reference are reported, and they are not
equivalent. The first is a pretrained CGCNN~\cite{xie2018cgcnn} evaluated
directly on the holdout, using the formation-energy checkpoint distributed
with the original CGCNN repository. Because the holdout compositions were
almost certainly inside that model's Materials Project training corpus, its
error is a contaminated, optimistic bound and is labeled as such wherever it
appears (Table~\ref{tab:pretrained}).

The like-for-like comparison is a MEGNet~\cite{chen2019megnet} trained from
scratch on the identical grouped split, with the holdout labels excluded from
every stage. A single configuration was used, with no architecture or
hyperparameter search: node and edge embeddings of dimension 64, a
two-dimensional state embedding, three MEGNet blocks ($1.92\times10^{5}$
trainable parameters), Adam at learning rate $10^{-3}$, batch size 32, a
\SI{5.0}{\angstrom} graph cutoff, seed 42, and a fixed 80-epoch budget. The
first grouped fold of the training partition (217/55 structures for the
formation energy, 114/29 for the gap, trained on the non-metal subset) served
as a monitored validation set, but no early stopping was applied and the final
epoch's weights were used. One implementation detail deviates slightly from
strict holdout isolation: the element vocabulary of the graph encoder was
built from the training and holdout structures together (compositions only, no
labels), we note it for completeness. Results are in Sec.~\ref{sec:perf} and
Table~\ref{tab:pretrained}, the limits of this comparison are discussed in
Sec.~\ref{sec:limitations}.

\begin{table}[t]
\centering
\caption{Graph-network references on the primary holdout. The pretrained
CGCNN figure (formation-energy checkpoint from the original repository) is a
contaminated, optimistic bound: the holdout compositions were almost certainly
in its Materials Project training corpus. The from-scratch MEGNet rows are the
like-for-like comparison, trained on the identical grouped split with the
holdout labels excluded (single untuned configuration, see
Secs.~\ref{sec:gnn} and \ref{sec:limitations}).}
\label{tab:pretrained}
\footnotesize
\begin{tabular}{@{}l l c c c@{}}
\toprule
Model & Target & $N$ & MAE & $R^2$ \\
\midrule
CatBoost (this work) & $\Eform$ & 48 & 0.0865 & 0.977 \\
CGCNN (pretrained, contaminated) & $\Eform$ & 48 & 0.1254 & 0.953 \\
MEGNet (from scratch, same split) & $\Eform$ & 48 & 0.2085 & 0.888 \\
\midrule
CatBoost (this work) & $\Eg$ & 19 & 0.6447 & $-0.069$ \\
MEGNet (from scratch, same split) & $\Eg$ & 19 & 0.7520 & $-0.190$ \\
\bottomrule
\end{tabular}
\end{table}

\begin{table}[t]
\centering
\caption{Dataset provenance and evaluation configuration. The LOCO row lists
the nine-cation extrapolation probe set (dominant-site occurrence $\ge 10$),
not the full cation content of the dataset, which also includes
lanthanide-bearing chalcogenides.}
\label{tab:dataset}
\footnotesize
\begin{tabular}{@{}ll@{}}
\toprule
Quantity & Value \\
\midrule
Source & Materials Project, space group~227 \\
Stoichiometries admitted & $AB_2X_4$ and single-cation $A_3X_4$ \\
Tetragonal branch (sg.~141) & Excluded \\
Curated entries & 320 \\
Training / holdout & 272 / 48 (grouped, seed 42) \\
Shared reduced formulas & 0 \\
Rejected: stoichiometry filter & 1059 \\
Rejected: descriptor failure & 62 \\
Rejected: missing $\Eform$ & 0 \\
Anion classes observed & N, O, S, Se \\
LOCO probe cations & Al, Co, Cr, Cu, Fe, Mg, Mn, Ni, Zn \\
Metallicity label & $\Eg\le\SI{0.01}{eV}$ (where gap reported) \\
Raw descriptors & 264 (120 Magpie, 19 spinel, \\
                & \quad 3 occupancy block, 122 CrystalNN) \\
Excluded from features & $\Ehull$, density, volume, \texttt{is\_stable}, \\
                       & \quad \texttt{SpaceGroupNumber}, \texttt{MendeleevNumber} \\
Optuna & TPE, 50 trials, grouped 3-fold inner CV \\
OOF selection & 4 repeats $\times$ grouped 5-fold \\
Baselines & median dummy, most-frequent dummy, \\
          & \quad RidgeCV, logistic regression \\
\bottomrule
\end{tabular}
\end{table}

\section{Results and Discussion}

\subsection{Surrogate performance and error analysis}
\label{sec:perf}

\begin{figure*}[htbp]
\centering
\subfloat[]{\includegraphics[width=0.4\textwidth]{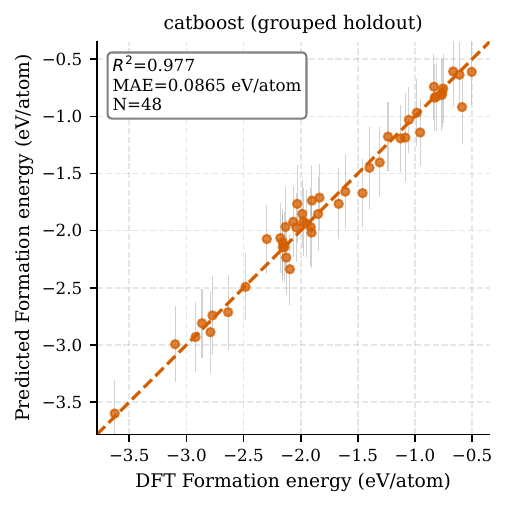}}\hfill
\subfloat[]{\includegraphics[width=0.4\textwidth]{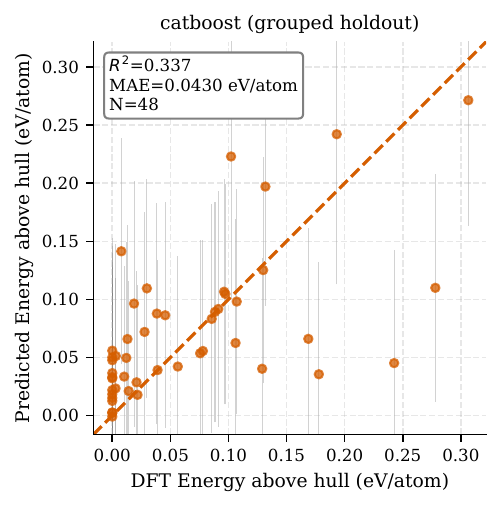}}\hfill
\subfloat[]{\includegraphics[width=0.4\textwidth]{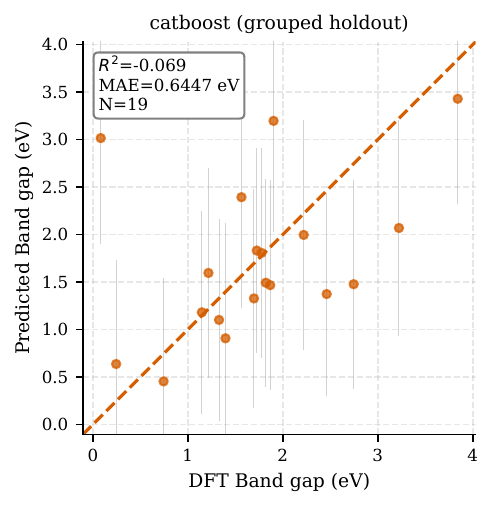}}\hfill
\subfloat[]{\includegraphics[width=0.4\textwidth]{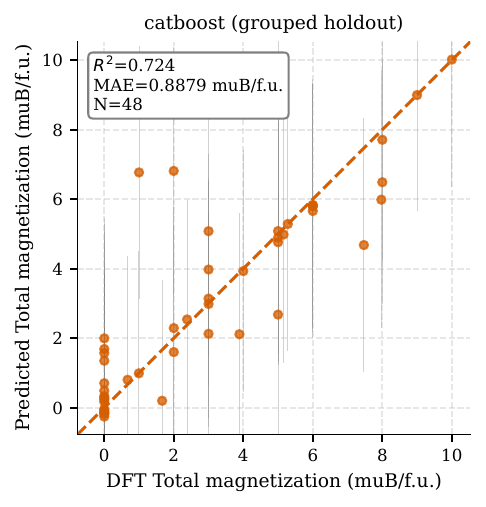}}
\caption{Predicted-versus-true parity plots on the primary holdout for the
regression champion (CatBoost), with the $y=x$ reference and per-point grouped
CV$+$ intervals (gray bars): (a)~formation energy, (b)~energy above hull,
(c)~band gap (non-metal subset), (d)~total magnetization. The gray intervals
are conservative for the formation energy (Sec.~\ref{sec:perf} and
Table~\ref{tab:conformal}) and should not be read as the point-error scale.}
\label{fig:parity}
\end{figure*}

\begin{figure*}[htbp]
\centering
\includegraphics[width=0.8\textwidth]{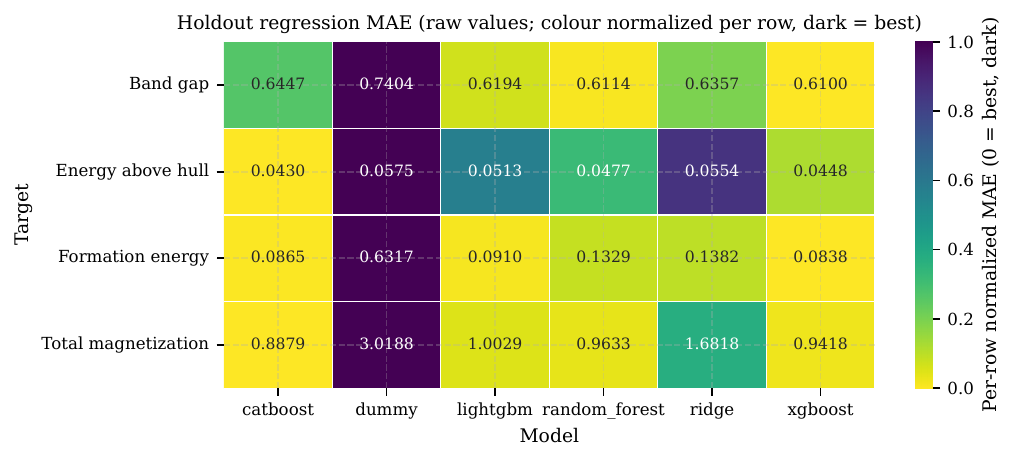}
\caption{Holdout regression MAE for every model and target (raw values
annotated, color normalized per target row, with the worst model in each row at
the dark end of a colorblind-safe scale). CatBoost and XGBoost are the
best-performing models across the energy and magnetization targets, no model
separates from the others on the band gap.}
\label{fig:heatmap}
\end{figure*}

\begin{figure*}[htbp]
\centering
\subfloat[]{\includegraphics[width=0.4\textwidth]{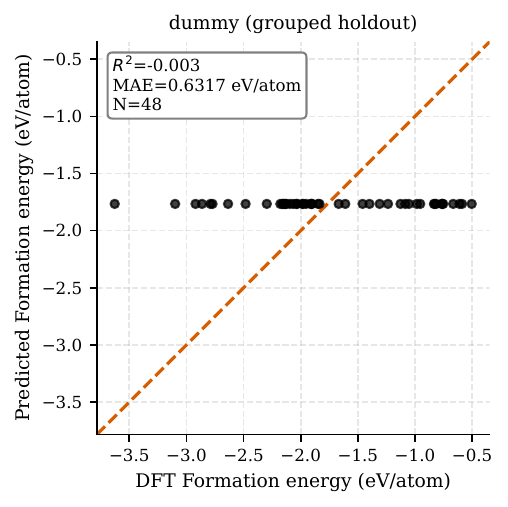}}
\subfloat[]{\includegraphics[width=0.4\textwidth]{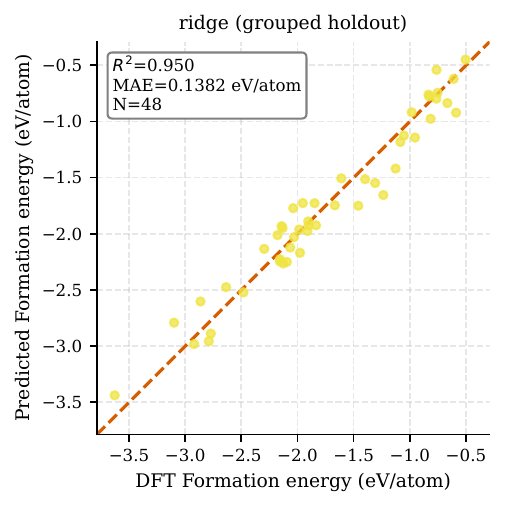}}\\
\subfloat[]{\includegraphics[width=0.4\textwidth]{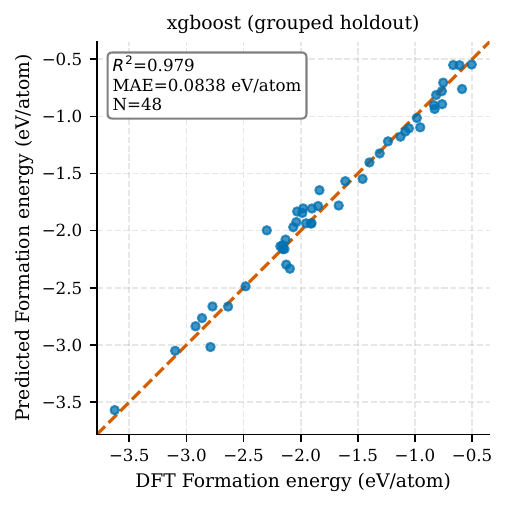}}
\subfloat[]{\includegraphics[width=0.4\textwidth]{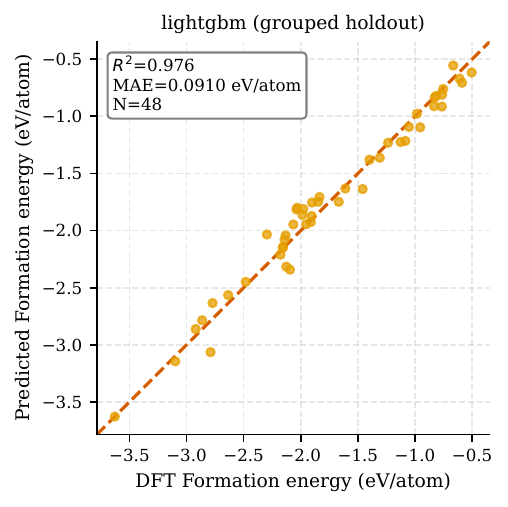}}\\
\subfloat[]{\includegraphics[width=0.4\textwidth]{fig_parity_formation_energy_per_atom_catboost.pdf}}
\subfloat[]{\includegraphics[width=0.4\textwidth]{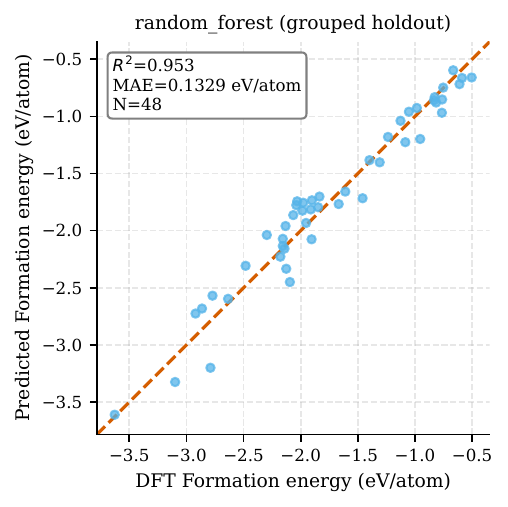}}
\caption{Per-model formation-energy parity plots on the holdout: (a)~dummy,
(b)~ridge, (c)~XGBoost, (d)~LightGBM, (e)~CatBoost champion,
(f)~RandomForest. The median baseline (a) collapses to a constant, the linear
baseline (b) captures the gross trend, and the boosted models (c-e) agree
tightly.}
\label{fig:parity_models}
\end{figure*}

\begin{table}[t]
\centering
\caption{Grouped CV$+$ conformal diagnostics (MAPIE, "plus" method over a
grouped five-fold split, $\alpha=0.10$, nominal 90\%) for the regression
champion of each target on the primary holdout: empirical coverage, its 95\%
binomial confidence interval, and mean interval width. Widths should be read
against the data scale: the formation-energy intervals over-cover
(coverage 1.000 at nominal 0.90) with a mean width of roughly seven times the
point MAE, a conservatism discussed in Sec.~\ref{sec:perf} and inherited by
the screening gate of Sec.~\ref{sec:screening}, the magnetization interval
(7.09~\muBfu) exceeds twice the dummy MAE, so individual intervals for that
target are close to uninformative even though coverage is honest.}
\label{tab:conformal}
\footnotesize
\begin{tabular}{@{}l c c c@{}}
\toprule
Target & Coverage & 95\% CI & Mean width \\
\midrule
Formation energy (\si{eV/atom}) & 1.000 & $[0.926,\,1.000]$ & 0.622 \\
Energy above hull (\si{eV/atom}) & 0.896 & $[0.773,\,0.965]$ & 0.196 \\
Band gap (\si{eV})              & 0.737 & $[0.488,\,0.909]$ & 2.237 \\
Total magnetization (\muBfu)    & 0.958 & $[0.857,\,0.995]$ & 7.093 \\
\bottomrule
\end{tabular}
\end{table}

\begin{table*}[t]
\centering
\caption{Primary-holdout regression performance for all targets and models
(three-seed prediction ensembles, the OOF-selected champion is in boldface).
$N$ is the holdout size, the band gap is evaluated on the non-metal subset
only. Conformal diagnostics are in Table~\ref{tab:conformal}, repeated-holdout
statistics, which we treat as the headline generalization figures, are in
Table~\ref{tab:robustness}.}
\label{tab:holdout}
\footnotesize
\begin{threeparttable}
\begin{tabular}{@{}ll r r r r@{}}
\toprule
Target & Model & $N$ & MAE & RMSE & $R^2$ \\
\midrule
\multirow{6}{*}{Formation energy (\si{eV/atom})}
 & dummy          & 48 & 0.6317 & 0.7560 & $-0.0026$ \\
 & ridge          & 48 & 0.1382 & 0.1687 & $0.9501$ \\
 & xgboost        & 48 & 0.0838 & 0.1085 & $0.9793$ \\
 & lightgbm       & 48 & 0.0910 & 0.1177 & $0.9757$ \\
 & \textbf{catboost} & 48 & \textbf{0.0865} & \textbf{0.1145} & $\mathbf{0.9770}$ \\
 & random\_forest & 48 & 0.1329 & 0.1636 & $0.9530$ \\
\midrule
\multirow{6}{*}{Energy above hull (\si{eV/atom})}
 & dummy          & 48 & 0.0575 & 0.0807 & $-0.1010$ \\
 & ridge          & 48 & 0.0554 & 0.0772 & $-0.0072$ \\
 & xgboost        & 48 & 0.0448 & 0.0669 & $0.2445$ \\
 & lightgbm       & 48 & 0.0513 & 0.0733 & $0.0910$ \\
 & \textbf{catboost} & 48 & \textbf{0.0430} & \textbf{0.0626} & $\mathbf{0.3372}$ \\
 & random\_forest & 48 & 0.0477 & 0.0691 & $0.1919$ \\
\midrule
\multirow{6}{*}{Band gap (\si{eV})}
 & dummy          & 19 & 0.7404 & 0.9810 & $-0.1910$ \\
 & ridge          & 19 & 0.6357 & 0.8647 & $0.0745$ \\
 & xgboost        & 19 & 0.6100 & 0.8996 & $-0.0016$ \\
 & lightgbm       & 19 & 0.6194 & 0.9137 & $-0.0334$ \\
 & \textbf{catboost} & 19 & \textbf{0.6447} & \textbf{0.9292} & $\mathbf{-0.0686}$ \\
 & random\_forest & 19 & 0.6114 & 0.8933 & $0.0124$ \\
\midrule
\multirow{6}{*}{Total magnetization (\muBfu)}
 & dummy          & 48 & 3.0188 & 4.1843 & $-1.0502$ \\
 & ridge          & 48 & 1.6818 & 2.0438 & $0.5109$ \\
 & xgboost        & 48 & 0.9418 & 1.8250 & $0.6100$ \\
 & lightgbm       & 48 & 1.0029 & 1.7948 & $0.6228$ \\
 & \textbf{catboost} & 48 & \textbf{0.8879} & \textbf{1.5343} & $\mathbf{0.7243}$ \\
 & random\_forest & 48 & 0.9633 & 1.7379 & $0.6463$ \\
\bottomrule
\end{tabular}
\begin{tablenotes}[flush]\footnotesize
\item $R^2$ is computed relative to the holdout mean, negative values indicate
a model worse than predicting that mean. The dummy baseline predicts the
training-fold median.
\end{tablenotes}
\end{threeparttable}
\end{table*}

\begin{figure*}[htbp]
\centering
\subfloat[]{\includegraphics[width=0.4\textwidth]{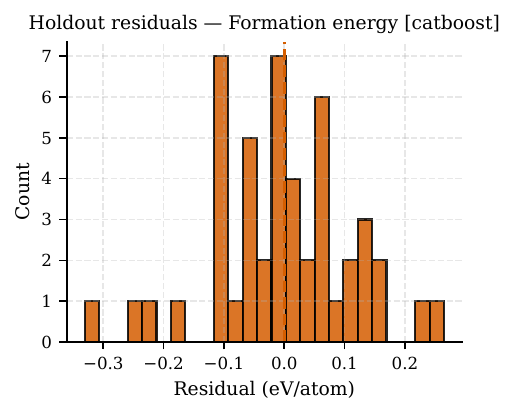}}\hfill
\subfloat[]{\includegraphics[width=0.4\textwidth]{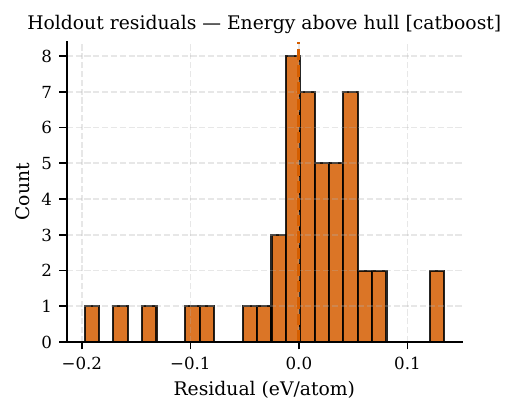}}\hfill
\subfloat[]{\includegraphics[width=0.4\textwidth]{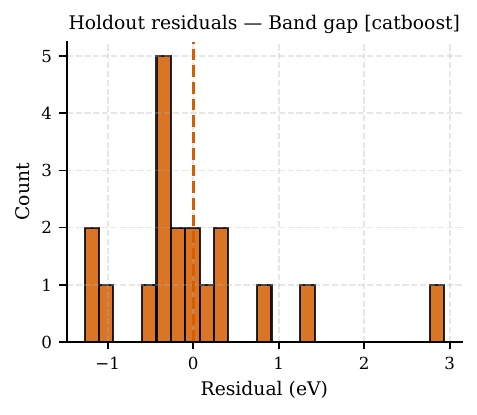}}\hfill
\subfloat[]{\includegraphics[width=0.4\textwidth]{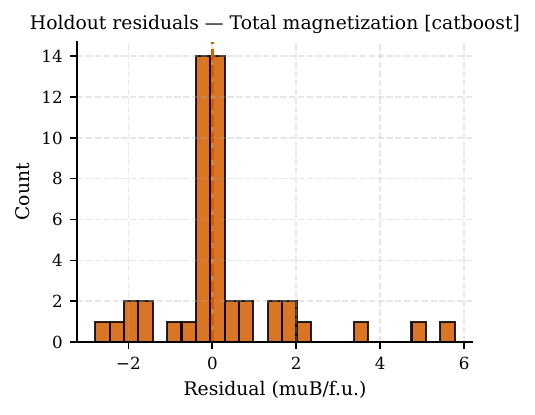}}
\caption{Holdout residual distributions for the champion of each target:
(a)~formation energy, (b)~energy above hull, (c)~band gap, (d)~total
magnetization. The energy residuals are peaked near zero, the magnetization
residuals show heavier tails, consistent with the wide conformal width in
Table~\ref{tab:conformal}.}
\label{fig:residuals}
\end{figure*}

Table~\ref{tab:holdout} reports the primary-holdout regression results for
every model, with the OOF-selected champion in boldface, Table~\ref{tab:metal}
the metallicity results, Table~\ref{tab:oof} the OOF selection statistics with
the paired-bootstrap tests, and Table~\ref{tab:conformal} the conformal
diagnostics. The OOF-selected champion was CatBoost for the four regression
targets and LightGBM for metallicity. Champion parity plots are shown in
Fig.~\ref{fig:parity}, residual distributions in Fig.~\ref{fig:residuals},
and the model$\times$target MAE landscape in Fig.~\ref{fig:heatmap}.

\paragraph{Formation energy.}
On the primary holdout the CatBoost champion reaches an MAE of
\SI{0.0865}{eV/atom} ($R^2=0.977$), against \SI{0.6317}{eV/atom} for the
median dummy and \SI{0.1382}{eV/atom} for ridge. Paired bootstrap testing
confirms the improvement over the dummy ($\Delta\mathrm{MAE}=-0.5445$, 95\% CI
$[-0.666,-0.422]$) and over ridge ($-0.0517$, CI $[-0.082,-0.019]$), the
differences against XGBoost ($+0.0027$) and LightGBM ($-0.0045$) are not
significant, so several boosted ensembles are effectively interchangeable here
(per-model parity plots in Fig.~\ref{fig:parity_models}). Over the twenty
repeated holdouts the champion averages $0.1206\pm0.0301$~eV/atom
($R^2=0.922\pm0.068$), and we take this, not the single split, as the headline
generalization figure. Part of the gap between the two numbers is mechanical:
the primary-split predictions average three seed replicates while the repeated
splits use single fits (Sec.~\ref{sec:pipe}), the remainder is ordinary split
variance. For context, an error near 0.09-0.12~eV/atom is of the same order
as the discrepancy between DFT formation enthalpies and
experiment~\cite{kirklin2015}, adequate for ranking chemistries but not for
resolving fine competition between similar compositions.

\paragraph{Energy above hull.}
The champion attains an MAE of \SI{0.0430}{eV/atom} ($R^2=0.337$) on the
primary holdout, versus \SI{0.0575}{eV/atom} (dummy) and \SI{0.0554}{eV/atom}
(ridge), repeated holdouts give $0.0476\pm0.0133$~eV/atom
($R^2=0.42\pm0.22$). The improvements over dummy ($-0.0145$, CI
$[-0.028,-0.000]$), ridge ($-0.0124$, CI $[-0.022,-0.003]$), and LightGBM
($-0.0083$, CI $[-0.015,-0.002]$) are significant, the difference against
XGBoost is not. Two honest readings follow. The low $R^2$ partly reflects the
narrow dynamic range of hull distances in this curated, near-stable set, on
which variance-based metrics are punishing. But the MAE itself is comparable
to the \SIrange{0.025}{0.050}{eV/atom} scale usually quoted for
room-temperature polymorph competition, and an error the size of the decision
window means compounds near the stability boundary cannot be classified
reliably. Consistent with the general finding of Ref.~\cite{bartel2020}, we
therefore use the hull model only as a coarse filter with wide margins, and
Sec.~\ref{sec:screening} reports how strongly any shortlist depends on the
chosen thresholds. The residual histogram [Fig.~\ref{fig:residuals}(b)] is
peaked near zero with a thin tail of under-stabilized outliers.
\begin{table}[t]
\centering
\caption{Metallicity classification. (A)~Per-model primary-holdout
performance (the OOF-selected champion, LightGBM, in boldface). The
most-frequent dummy labels every holdout entry non-metallic, which is why its
accuracy (0.396) lies below the holdout majority class (29/48 = 0.604), the
informative naive reference. CatBoost ties LightGBM on every metric: the two
boosters' thresholded predictions coincide on this split (Sec.~\ref{sec:perf}).
Champion balanced accuracy and Matthews coefficient, derived from the
confusion matrix of Fig.~\ref{fig:confmat}, are 0.843 and 0.693.
(B)~Sensitivity of the champion to the band-gap threshold defining the label,
computed by grouped five-fold cross-validation on the 272-entry training
partition.}

\label{tab:metal}
\footnotesize
\begin{tabular}{@{}l r r r@{}}
\toprule
\multicolumn{4}{@{}l}{\textit{(A) Per-model holdout ($N=48$)}}\\
Model & Accuracy & F1 & AUC \\
\midrule
dummy (most frequent) & 0.3958 & 0.0000 & 0.5000 \\
ridge                 & 0.7917 & 0.8214 & 0.8421 \\
xgboost               & 0.8125 & 0.8421 & 0.9256 \\
\textbf{lightgbm}     & \textbf{0.8542} & \textbf{0.8814} & \textbf{0.9328} \\
catboost              & 0.8542 & 0.8814 & 0.9328 \\
random\_forest        & 0.8958 & 0.9153 & 0.9093 \\
\midrule
\multicolumn{4}{@{}l}{\textit{(B) Label-threshold sensitivity (training OOF)}}\\
Threshold (\si{eV}) & Accuracy & F1 & AUC \\
\midrule
0.01 & 0.8235 & 0.8125 & 0.9049 \\
0.05 & 0.8309 & 0.8231 & 0.9124 \\
0.10 & 0.8419 & 0.8377 & 0.9124 \\
\bottomrule
\end{tabular}
\end{table}

\begin{table*}[t]
\centering
\caption{Left columns: repeated grouped five-fold out-of-fold statistics
(mean $\pm$ sd over four repeats) on the training partition. Because
hyperparameters were tuned on the same partition without nesting, these are
model-selection scores, mildly optimistic as performance estimates
(Sec.~\ref{sec:pipe}). Right column: champion-versus-baseline MAE differences
on the \emph{primary holdout}, quoted as the point difference of the holdout
MAEs of Table~\ref{tab:holdout} (champion minus comparator), significance is
assessed by paired bootstrap (2000 resamples, $\ast$ marks a 95\% percentile
interval excluding zero, ns otherwise).}
\label{tab:oof}
\footnotesize
\begin{tabular}{@{}ll c c c l@{}}
\toprule
Target & Model & MAE (mean$\pm$sd) & RMSE (mean$\pm$sd) & $R^2$ / Acc.\ (mean$\pm$sd) & Holdout $\Delta$MAE \\
\midrule
\multirow{3}{*}{Formation energy}
 & dummy          & $0.6601\pm0.0018$ & $0.8496\pm0.0005$ & $-0.0070\pm0.0011$ & vs.\ dummy: $-0.5445$ [$\ast$] \\
 & ridge          & $0.2000\pm0.0045$ & $0.3359\pm0.0363$ & $0.8412\pm0.0338$ & vs.\ ridge: $-0.0517$ [$\ast$] \\
 & \textbf{catboost} & $\mathbf{1.322\times10^{-1}\pm0.0043}$ & $\mathbf{0.2458\pm0.0145}$ & $\mathbf{0.9156\pm0.0098}$ & vs.\ xgboost: $+0.0027$ [ns] \\
\midrule
\multirow{3}{*}{Energy above hull}
 & dummy          & $0.0663\pm0.0003$ & $0.1324\pm0.0004$ & $-0.0781\pm0.0073$ & vs.\ dummy: $-0.0145$ [$\ast$] \\
 & ridge          & $0.0590\pm0.0018$ & $0.1024\pm0.0062$ & $0.3525\pm0.0779$ & vs.\ ridge: $-0.0124$ [$\ast$] \\
 & \textbf{catboost} & $\mathbf{0.0487\pm0.0008}$ & $\mathbf{0.1071\pm0.0011}$ & $\mathbf{0.2948\pm0.0150}$ & vs.\ lightgbm: $-0.0083$ [$\ast$] \\
\midrule
\multirow{3}{*}{Band gap}
 & dummy          & $0.9046\pm0.0021$ & $1.2349\pm0.0024$ & $-0.0792\pm0.0043$ & vs.\ dummy: $-0.0945$ [ns] \\
 & ridge          & $0.6610\pm0.0211$ & $0.9712\pm0.0432$ & $0.3314\pm0.0588$ & vs.\ ridge: $+0.0079$ [ns] \\
 & \textbf{catboost} & $\mathbf{0.5000\pm0.0341}$ & $\mathbf{0.7602\pm0.0378}$ & $\mathbf{0.5903\pm0.0399}$ & vs.\ xgboost: $+0.0353$ [ns] \\
\midrule
\multirow{3}{*}{Total magnetization}
& dummy          & $2.5376\pm0.0338$ & $4.2256\pm0.0093$ & $-0.5085\pm0.0067$ & vs.\ dummy: $-2.1382$ [$\ast$] \\
 & ridge          & $2.2127\pm0.0684$ & $2.9564\pm0.0736$ & $0.2613\pm0.0369$ & vs.\ ridge: $-0.7971$ [$\ast$] \\
 & \textbf{catboost} & $\mathbf{1.3084\pm0.0135}$ & $\mathbf{2.0047\pm0.0271}$ & $\mathbf{0.6604\pm0.0092}$ & vs.\ lightgbm: $-0.1128$ [ns] \\
\midrule
\multirow{3}{*}{Metallicity (Acc/F1/AUC)}
 & dummy            & \multicolumn{3}{c}{$0.5101\pm0.0312$ / $0.0516\pm0.1032$ / $0.4870\pm0.0261$} & - \\
 & ridge         & \multicolumn{3}{c}{$0.7390\pm0.0124$ / $0.7183\pm0.0110$ / $0.8145\pm0.0166$} & - \\
 & \textbf{lightgbm} & \multicolumn{3}{c}{$\mathbf{0.8198\pm0.0161}$ / $\mathbf{0.8124\pm0.0193}$ / $\mathbf{0.9069\pm0.0105}$} & - \\
\bottomrule
\end{tabular}
\end{table*}

\paragraph{Band gap.}
On the 19-member non-metal holdout subset the champion records an MAE of
\SI{0.6447}{eV} and $R^2=-0.069$, and the paired bootstrap shows no
significant advantage over the dummy baseline ($\Delta\mathrm{MAE}=-0.0945$,
CI $[-0.439,0.219]$) or over any other model. We report this as a negative
result: on the present non-metal subset, no model demonstrably beats the
trivial baseline for gap regression. Two causes are plausible. The PBE-level
labels compress and distort the target, most severely for Mott and
charge-transfer oxides (Sec.~\ref{sec:limitations}), and the evaluation subset
is small, which the wide conformal coverage interval for this target reflects
(Table~\ref{tab:conformal}). Cross-validated scores on the training partition
are considerably better ($R^2\approx0.59$, Table~\ref{tab:oof}), but those
scores share data with the tuning loop and are optimistic
(Sec.~\ref{sec:pipe}), the more reliable indication is the learning curve
below, which shows the target is still data-limited at the largest training
size sampled.

\paragraph{Total magnetization.}
The champion reaches an MAE of $0.8879$~\muBfu{} ($R^2=0.724$) on the
primary holdout, against 3.0188~\muBfu{} (dummy) and 1.6818~\muBfu{} (ridge);
repeated holdouts give $1.273\pm0.193$~\muBfu{} ($R^2=0.61\pm0.13$), again the
figure we treat as the headline. The champion beats the dummy ($-2.1382$, CI
$[-3.051,-1.235]$) and ridge ($-0.7971$, CI $[-1.130,-0.426]$), differences
against the other ensembles are not significant. The conformal interval for
this target is very wide (mean width 7.09~\muBfu{}, Table~\ref{tab:conformal}),
wider than the naive spread of the data itself, so although the point accuracy
is useful, individual predictions carry little decision value on their own.

\begin{figure*}[htbp]
\centering
\subfloat[]{\includegraphics[width=0.32\textwidth]{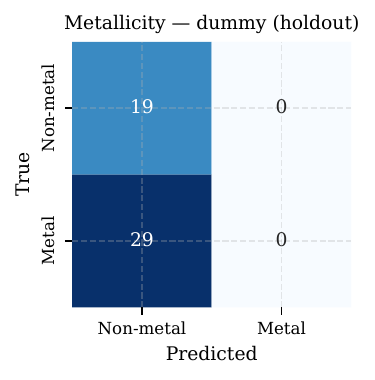}}\hfill
\subfloat[]{\includegraphics[width=0.32\textwidth]{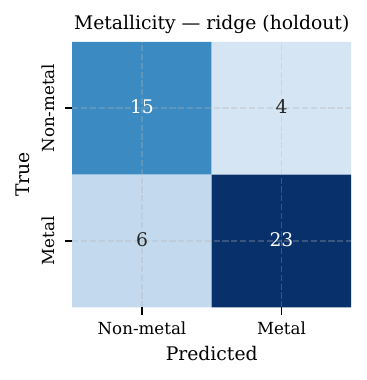}}\hfill
\subfloat[]{\includegraphics[width=0.32\textwidth]{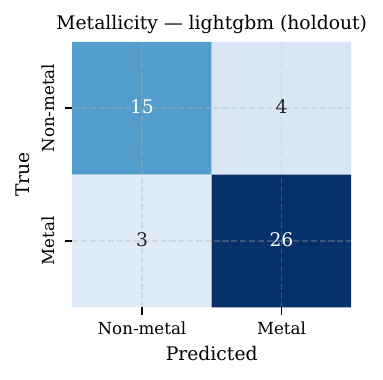}}\\
\subfloat[]{\includegraphics[width=0.32\textwidth]{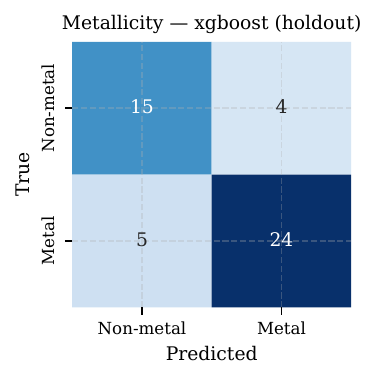}}\hfill
\subfloat[]{\includegraphics[width=0.32\textwidth]{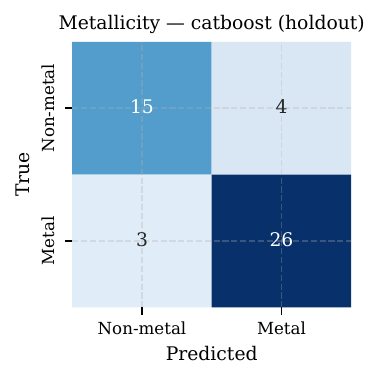}}\hfill
\subfloat[]{\includegraphics[width=0.32\textwidth]{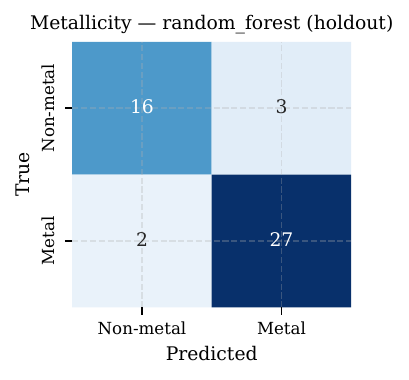}}
\caption{Holdout metallicity confusion matrices: (a)~dummy, (b)~ridge,
(c)~LightGBM champion, (d)~XGBoost, (e)~CatBoost, (f)~RandomForest. The
champion yields 15/4/3/26 (TN/FP/FN/TP), i.e.\ accuracy 0.8542, balanced
accuracy 0.843, MCC 0.693, CatBoost (e) coincides with the champion after
thresholding (Sec.~\ref{sec:perf}).}
\label{fig:confmat}
\end{figure*}

\paragraph{Metallicity.}
The LightGBM champion reaches a holdout accuracy of 0.8542, F1 of 0.8814
(metallic class positive), and AUC of 0.9328, the confusion matrix
[Fig.~\ref{fig:confmat}] is 15 true non-metals, 4 false metals, 3 false
non-metals, 26 true metals, from which the balanced accuracy is 0.843 and the
Matthews correlation coefficient 0.693. CatBoost ties the champion on every
holdout metric, its thresholded predictions coinciding entry for entry with
LightGBM's on this 48-entry split [Figs.~\ref{fig:confmat}(c) and (e)], ties
of this kind between closely related gradient boosters are plausible on a
small holdout, and the coincidence occurs after thresholding, not at the level
of the underlying probability scores.

The baselines need a comment. The most-frequent dummy classifier labels
everything non-metallic, because non-metals are the majority class in the
training partition, and scores 19/48~$=$~0.396 with $F1=0$ on the holdout,
where 29 of 48 entries are metallic, the informative naive reference on the
holdout is therefore the majority class at 0.604, and the train/holdout
class-balance shift itself illustrates the sampling noise of a 48-entry split.
The ridge baseline reaches 0.7917 (AUC 0.8421). RandomForest posts the best
single-split accuracy (0.8958) with a lower AUC (0.9093) than LightGBM and
CatBoost (0.9328), since selection was fixed on OOF AUC before the holdout was
seen, LightGBM remains the champion, and the RandomForest advantage is within
the split variance of Table~\ref{tab:robustness}. Over the repeated holdouts
the champion averages an accuracy of $0.847\pm0.058$.

\paragraph{Conformal calibration.}
The grouped CV$+$ intervals are honest but not uniformly tight
(Table~\ref{tab:conformal}). For the formation energy the empirical coverage
is 1.000 against the nominal 0.90, with a mean width of \SI{0.622}{eV/atom},
roughly seven times the point MAE: the intervals over-cover, because grouping
the calibration folds by reduced formula makes the residual quantiles
conservative wherever between-group heterogeneity exceeds within-group
scatter. This conservatism propagates into the screening gate of
Sec.~\ref{sec:screening}, where the formation-energy half-width is one of the
four criteria, the stringency of the width threshold there reflects the
calibration as much as the model. The hull intervals are close to nominal
(coverage 0.896), the magnetization intervals cover well (0.958) but are too
wide to be decision-useful, and the band-gap coverage (0.737, CI
$[0.488,0.909]$) undershoots nominal on the 19-entry subset, consistent with
the small-sample reading above.

\begin{figure*}[htbp]
\centering
\subfloat[]{\includegraphics[width=0.32\textwidth]{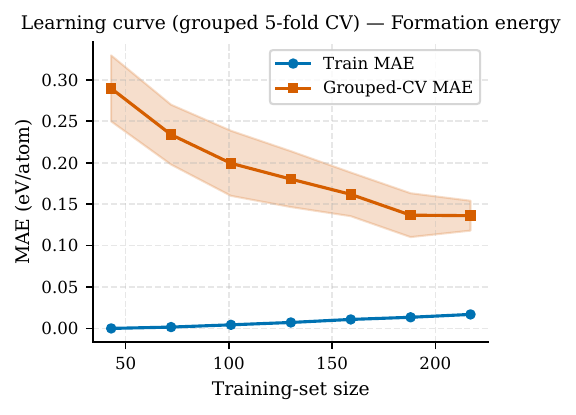}}\hfill
\subfloat[]{\includegraphics[width=0.32\textwidth]{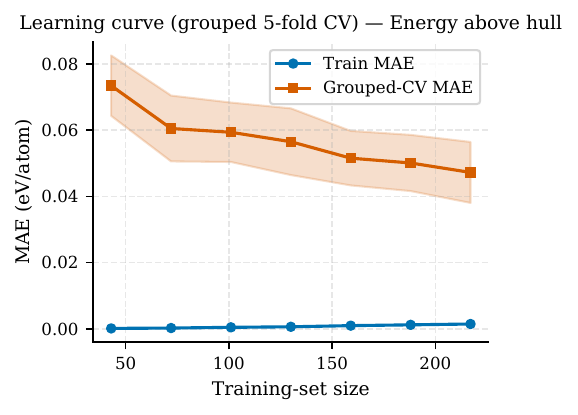}}\hfill
\subfloat[]{\includegraphics[width=0.32\textwidth]{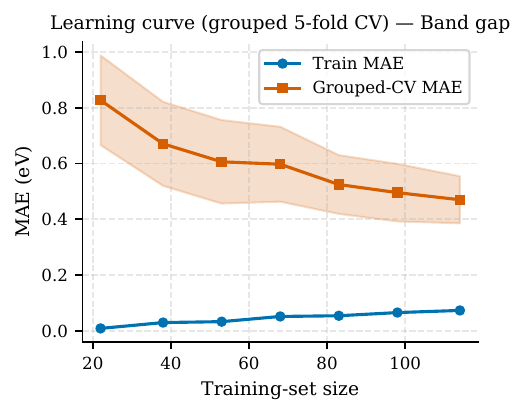}}
\caption{Grouped five-fold learning curves (train and CV MAE versus
training-set size): (a)~formation energy, (b)~energy above hull, (c)~band gap.
The energy targets reach a data-limited plateau by $N\approx190$-220, the
band-gap CV-MAE is still declining at the largest size sampled.}
\label{fig:learning}
\end{figure*}

\paragraph{Repeated holdouts and data efficiency.}
The twenty repeated splits (Table~\ref{tab:robustness}) give champion averages
of $0.1206\pm0.0301$~eV/atom (formation energy), $0.0476\pm0.0133$~eV/atom
(hull), $0.4608\pm0.1091$~eV (gap), $1.2732\pm0.1930$~\muBfu{}
(magnetization), and $0.8469\pm0.0575$ (metallicity accuracy). Grouped
five-fold learning curves (Fig.~\ref{fig:learning}) show the formation-energy
CV-MAE falling from \SI{0.290}{eV/atom} at $N=43$ to a plateau near
\SI{0.137}{eV/atom} by $N\approx188$-217, and the hull CV-MAE plateauing near
\SI{0.047}{eV/atom}, both energy targets are close to their data-limited
asymptote at the present sample size. The band-gap CV-MAE is still declining
at the largest size sampled (\SI{0.470}{eV} at $N=114$), which is what a
data-starved target looks like, and it is why we read the holdout gap result
as power-limited rather than as a ceiling.

\paragraph{Zero-inflation and threshold sensitivity.}
A hurdle treatment of the magnetization-a zero/nonzero classifier
($|m|>0.1$~\muBfu{}) composed with a magnitude regressor on the nonzero
subset-gives an OOF MAE of 1.2822~\muBfu{} with $R^2=0.448$, against
1.2938~\muBfu{} and $R^2=0.670$ for the direct CatBoost regression under the
same protocol (first OOF repetition in both cases). The hurdle MAE is
marginally lower, but its variance-weighted accuracy is much worse, and both
figures come from a single repetition, there is no consistent improvement to
justify the extra moving part, so we keep the direct model and report the
hurdle result as a null. The metallicity label threshold was varied over 0.01,
0.05, and \SI{0.10}{eV} in a grouped five-fold cross-validation on the
272-entry training partition (not on the holdout): accuracy rises from 0.8235
to 0.8309 to 0.8419 with AUC stable near 0.91 [Table~\ref{tab:metal}(B)], so
the residual error is dominated by near-zero-gap cases at the
metal-insulator boundary, and a threshold near the PBE numerical-noise floor
(\SI{0.10}{eV}) is marginally more robust.

\begin{table*}[t]
\centering
\caption{Robustness summary. Repeated-holdout statistics are the mean $\pm$
sd over 20 grouped splits of the full dataset, single-seed champion fits with
hyperparameters reused from the primary training partition
(Sec.~\ref{sec:robust}). LOAO/LOCO columns give the $R^2$ range over held-out
anion/cation chemistries (the corresponding per-cation MAE map is
Fig.~\ref{fig:loco}), the $y$-scrambled column the permutation-null $R^2$.
$N_{\mathrm{a}}$/$N_{\mathrm{c}}$ are the numbers of probed anion/cation
chemistries.}
\label{tab:robustness}
\footnotesize
\begin{tabular}{@{}l l c c c@{}}
\toprule
Target & Repeated holdout ($n{=}20$) & LOAO $R^2$ range ($N_{\mathrm a}{=}4$) & LOCO $R^2$ range & $y$-scrambled $R^2$ \\
\midrule
Formation energy & MAE $0.1206\pm0.0301$, $R^2\,0.9224\pm0.0680$ & $-1.560$ to $0.870$ & $0.614$ to $0.965$ ($N_{\mathrm c}{=}9$) & $-0.2072\pm0.1024$ \\
Energy above hull & MAE $0.0476\pm0.0133$, $R^2\,0.4221\pm0.2237$ & $-0.162$ to $0.613$ & $-0.381$ to $0.496$ ($N_{\mathrm c}{=}9$) & $-0.1471\pm0.0227$ \\
Band gap & MAE $0.4608\pm0.1091$, $R^2\,0.5348\pm0.1906$ & $-0.229$ to $0.018$ & $-1.956$ to $0.597$ ($N_{\mathrm c}{=}7$) & $-0.2516\pm0.1399$ \\
Total magnetization & MAE $1.2732\pm0.1930$, $R^2\,0.6072\pm0.1255$ & $-1.304$ to $0.626$ & $-1.630$ to $0.803$ ($N_{\mathrm c}{=}9$) & $-0.2341\pm0.0970$ \\
Metallicity (Acc/F1) & Acc $0.8469\pm0.0575$, F1 $0.8440\pm0.0588$ & n/a & n/a & n/a \\
\bottomrule
\end{tabular}
\end{table*}

\subsection{Feature attribution and descriptor ablation}
\label{sec:shap}

\begin{figure*}[htbp]
\centering
\subfloat[]{\includegraphics[width=0.45\textwidth]{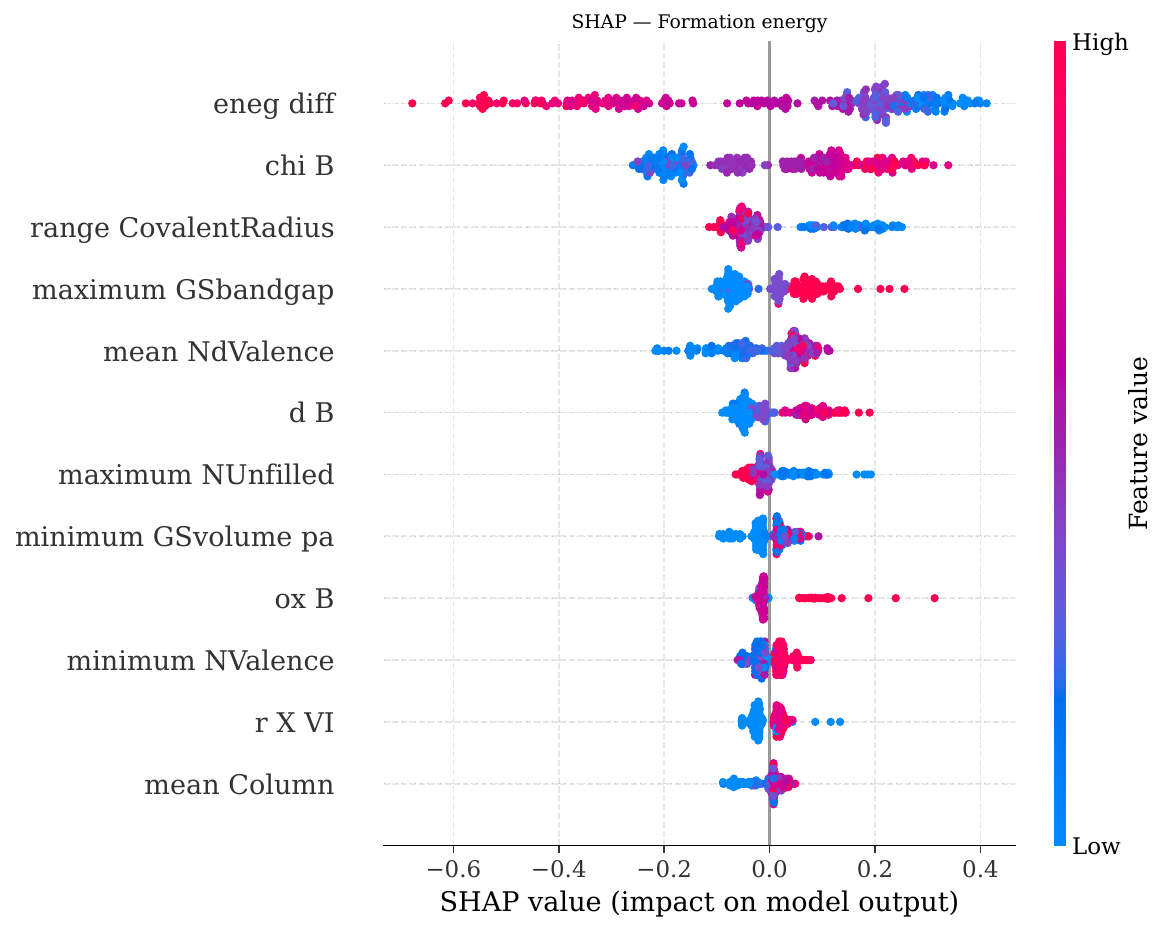}}\hfill
\subfloat[]{\includegraphics[width=0.45\textwidth]{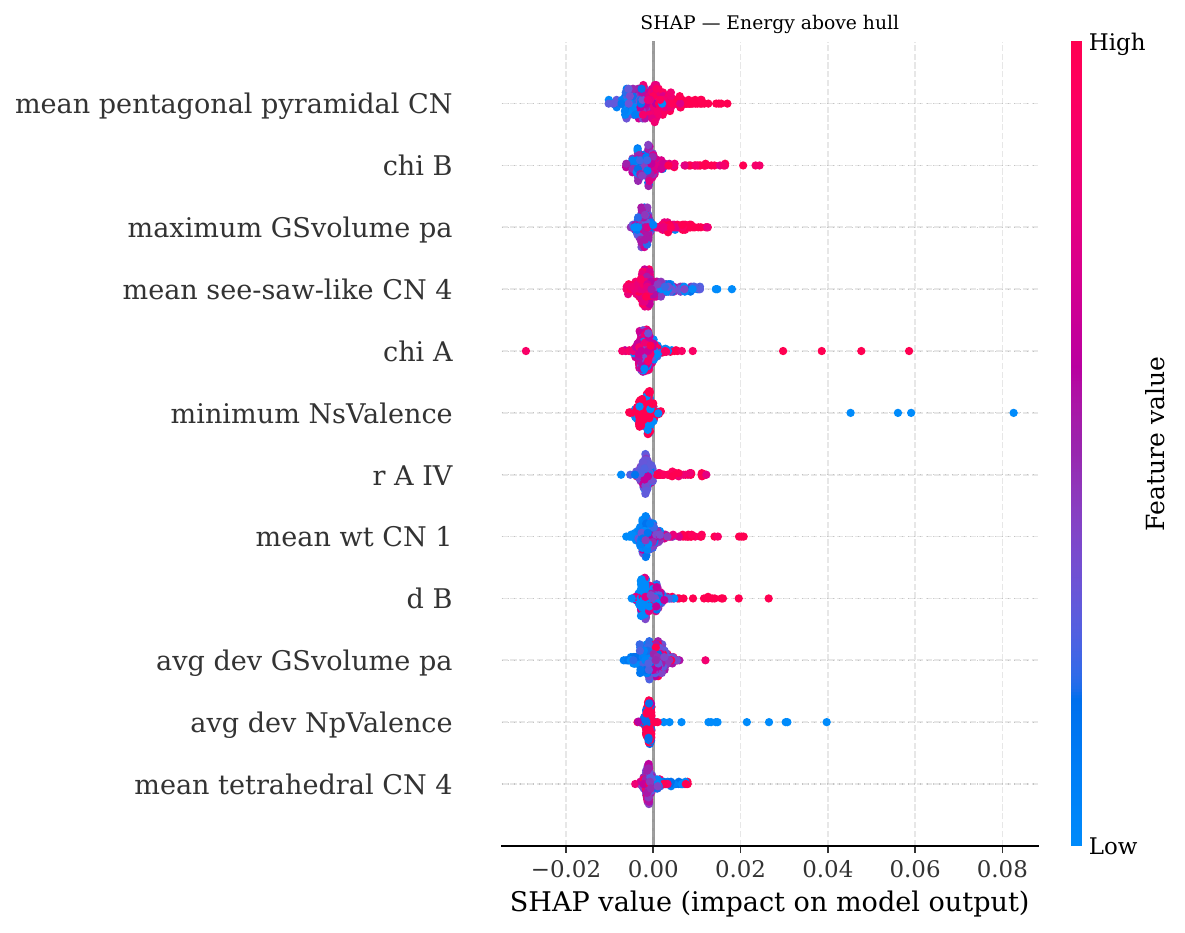}}\hfill
\subfloat[]{\includegraphics[width=0.45\textwidth]{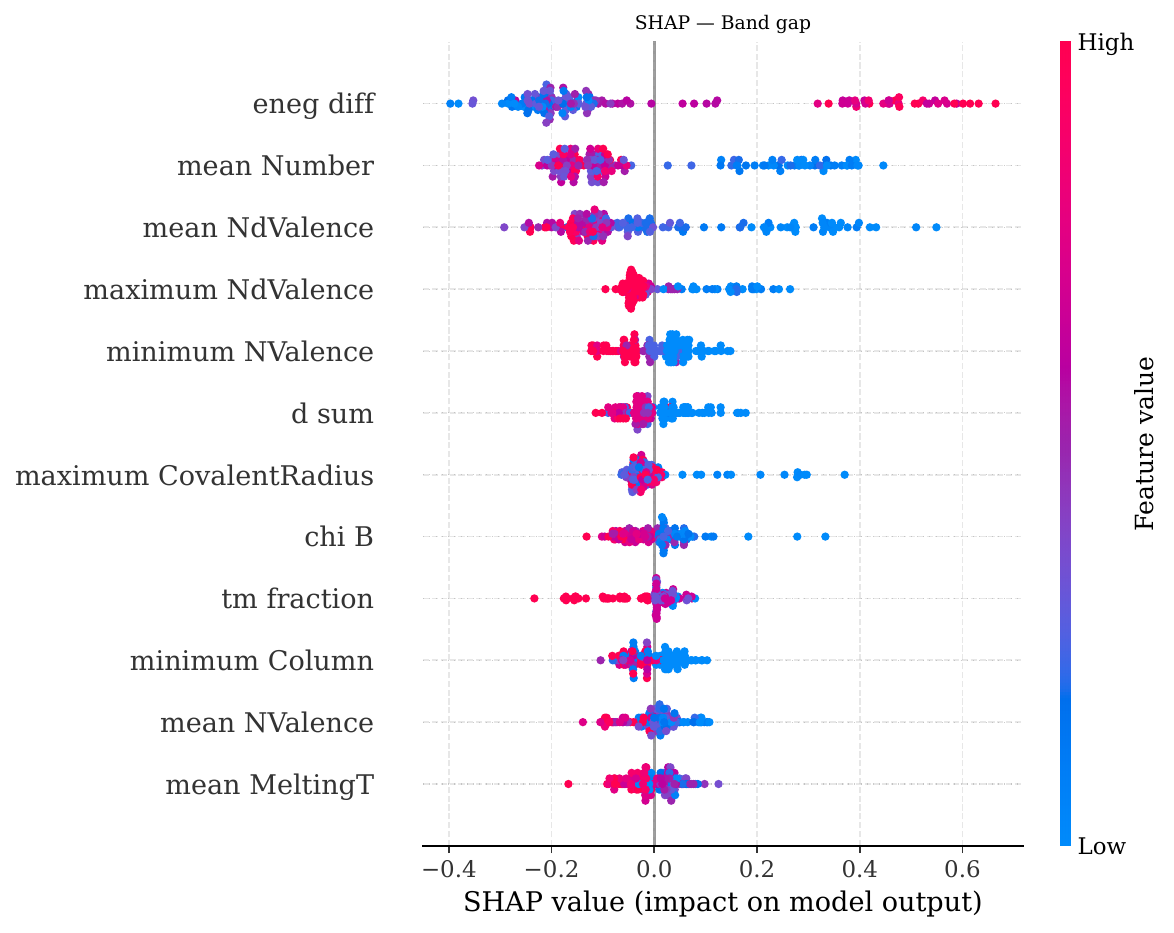}}\hfill
\subfloat[]{\includegraphics[width=0.45\textwidth]{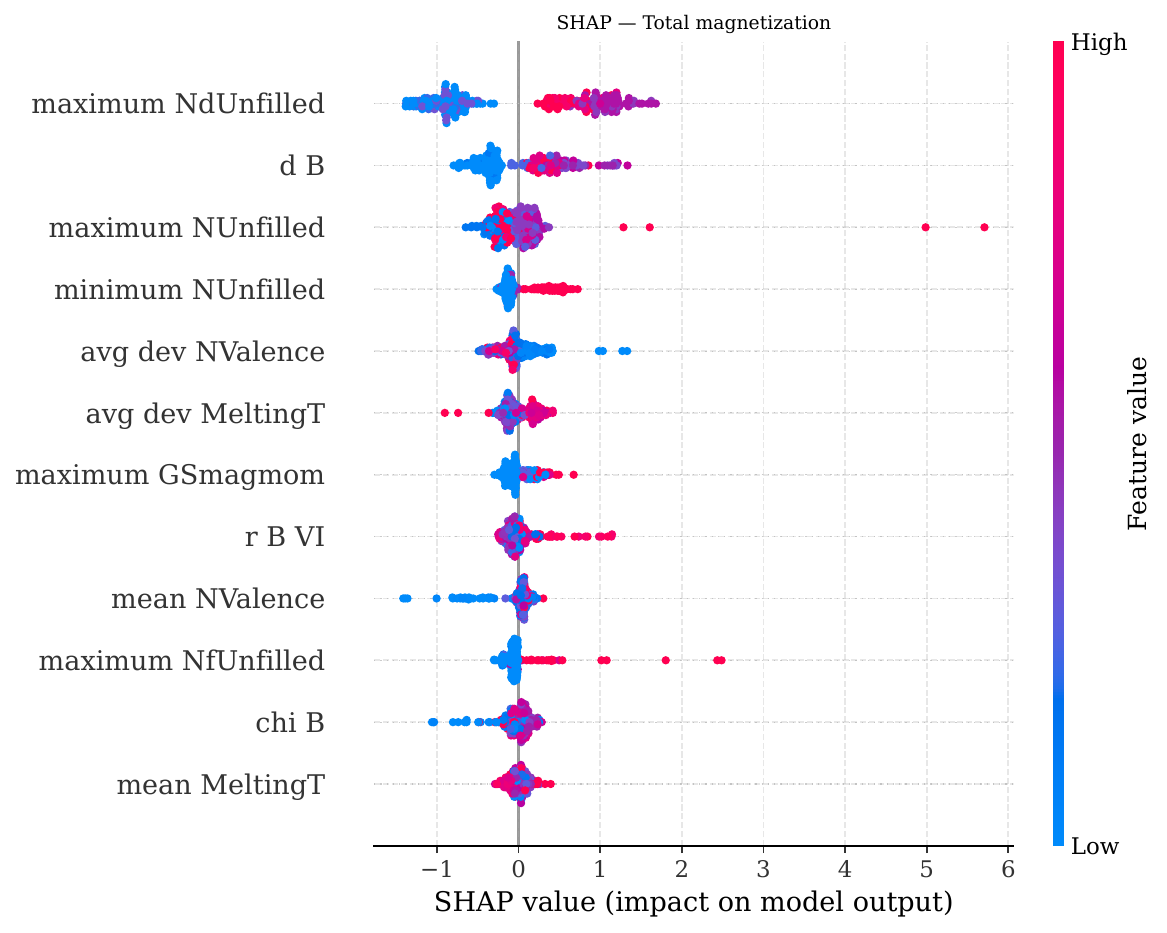}}
\caption{SHAP beeswarm summaries (grouped-OOF protocol, color encodes feature
value) for the champion of each target: (a)~formation energy, (b)~energy above
hull, (c)~band gap, (d)~total magnetization. For the formation energy, the
electronegativity descriptors $\Delta\chi$ and $\chi_B$ lead the ranking
(Sec.~\ref{sec:shap}). For the magnetization, the moment grows monotonically
with unfilled-$d$ occupancy and $d_B$, the localized-moment mechanism.}
\label{fig:beeswarm}
\end{figure*}

\begin{table}[t]
\centering
\caption{Leading SHAP features (mean $|$SHAP$|$, top six of the archived
top-15 lists) for the energy-above-hull, band-gap, and total-magnetization
champions. The formation-energy ranking is shown in
Figs.~\ref{fig:beeswarm}(a) and \ref{fig:shapbar}(a) and archived numerically
in the result bundle. Coordination-motif features originate from the CrystalNN
site fingerprints.}
\label{tab:shap}
\scriptsize
\begin{tabular}{@{}l l r@{}}
\toprule
Target & Feature & mean $|$SHAP$|$ \\
\midrule
\multirow{6}{*}{Band gap}
 & eneg\_diff & 0.2623 \\
 & MagpieData mean Number & 0.1833 \\
 & MagpieData mean NdValence & 0.1591 \\
 & MagpieData maximum NdValence & 0.0594 \\
 & $\chi_B$ & 0.0449 \\
 & tm\_fraction & 0.0400 \\
\midrule
\multirow{6}{*}{Energy above hull}
 & mean pentagonal-pyramidal $\mathrm{CN}_6$ & 0.00376 \\
 & $\chi_B$ & 0.00305 \\
 & MagpieData maximum GSvolume\_pa & 0.00302 \\
 & mean see-saw-like $\mathrm{CN}_4$ & 0.00286 \\
 & $\chi_A$ & 0.00281 \\
 & $d_B$ & 0.00225 \\
\midrule
\multirow{6}{*}{Total magnetization}
 & MagpieData maximum NdUnfilled & 0.8924 \\
 & $d_B$ & 0.4158 \\
 & MagpieData maximum NUnfilled & 0.2225 \\
 & MagpieData minimum NUnfilled & 0.2010 \\
 & $r_B^{VI}$ & 0.1268 \\
 & $\chi_B$ & 0.1146 \\
\bottomrule
\end{tabular}
\end{table}

\begin{figure}[t]
\centering
\includegraphics[width=\linewidth]{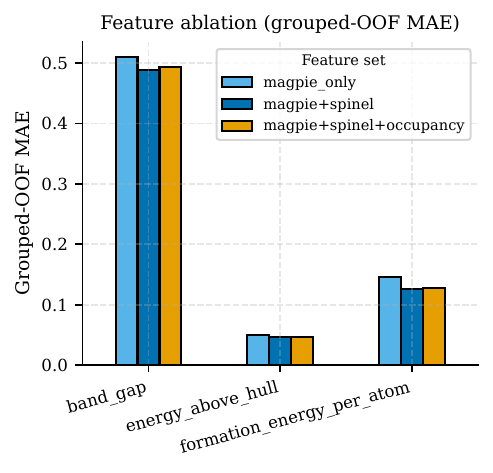}
\caption{Grouped-OOF MAE under the Magpie-only, Magpie$+$spinel, and
Magpie$+$spinel$+$occupancy feature sets. The bespoke spinel block lowers the
error for all three targets (largest relative gain for the formation energy), the
three-descriptor occupancy block adds nothing (Table~\ref{tab:ablation}).}
\label{fig:ablation}
\end{figure}

\begin{figure*}[htbp]
\centering
\subfloat[]{\includegraphics[width=0.48\textwidth]{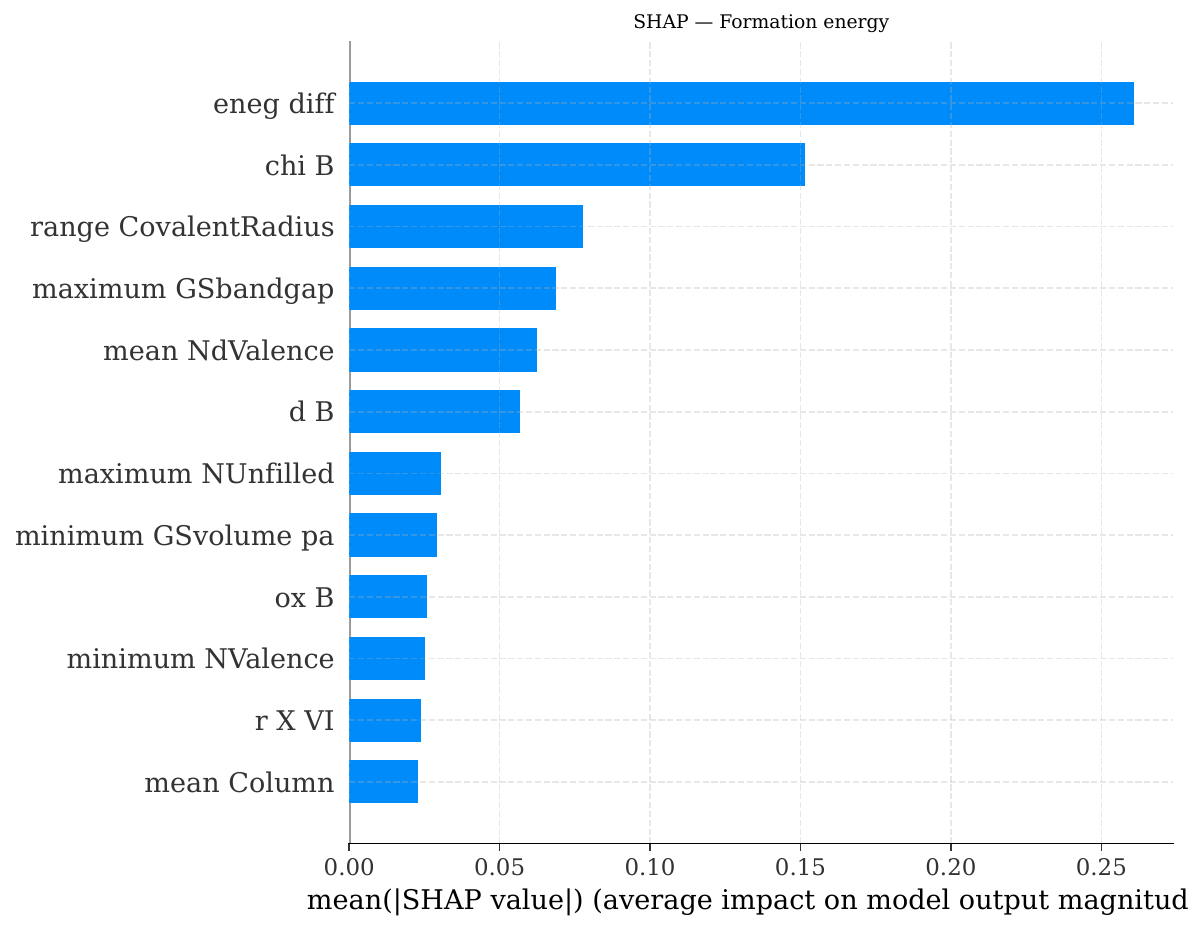}}\hfill
\subfloat[]{\includegraphics[width=0.48\textwidth]{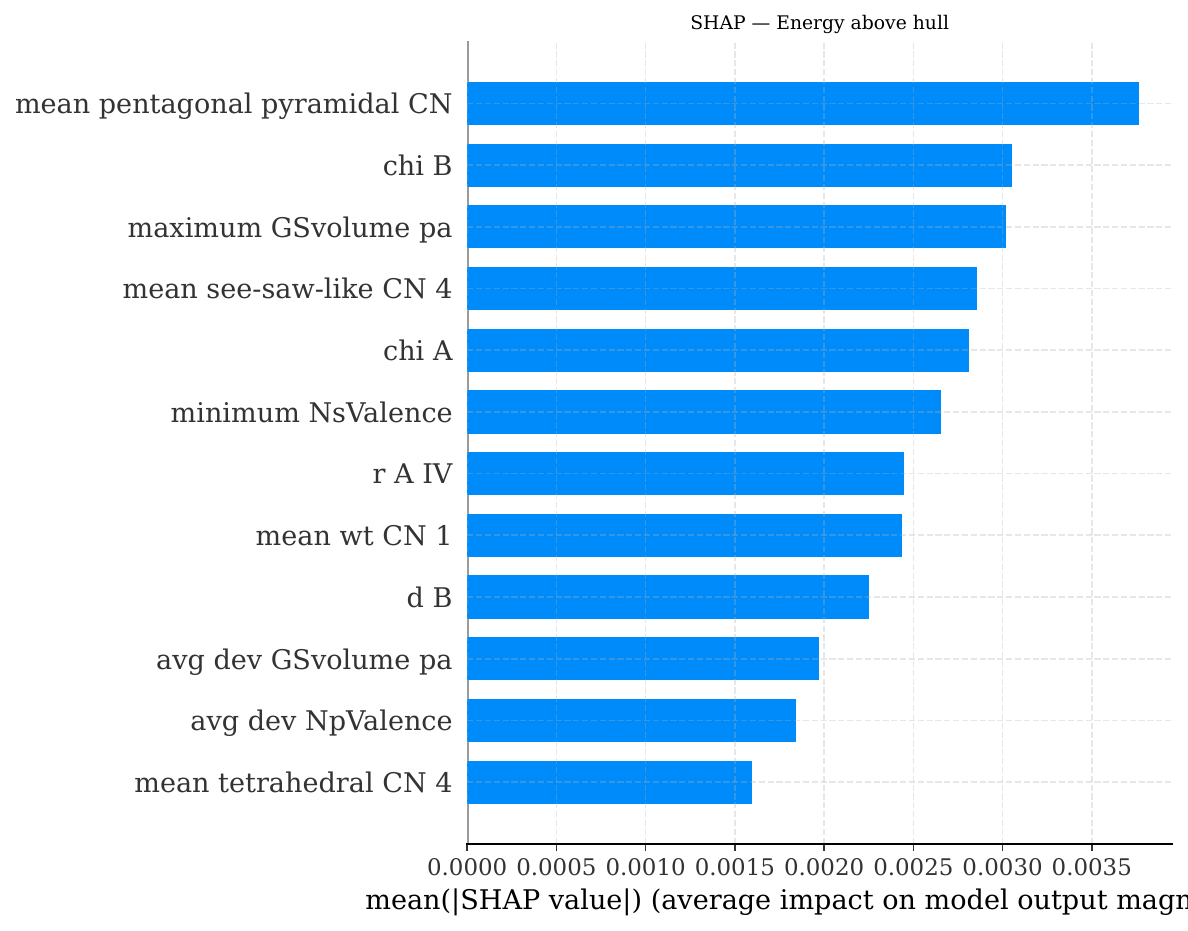}} \\ \vspace{0.2cm} 
\subfloat[]{\includegraphics[width=0.48\textwidth]{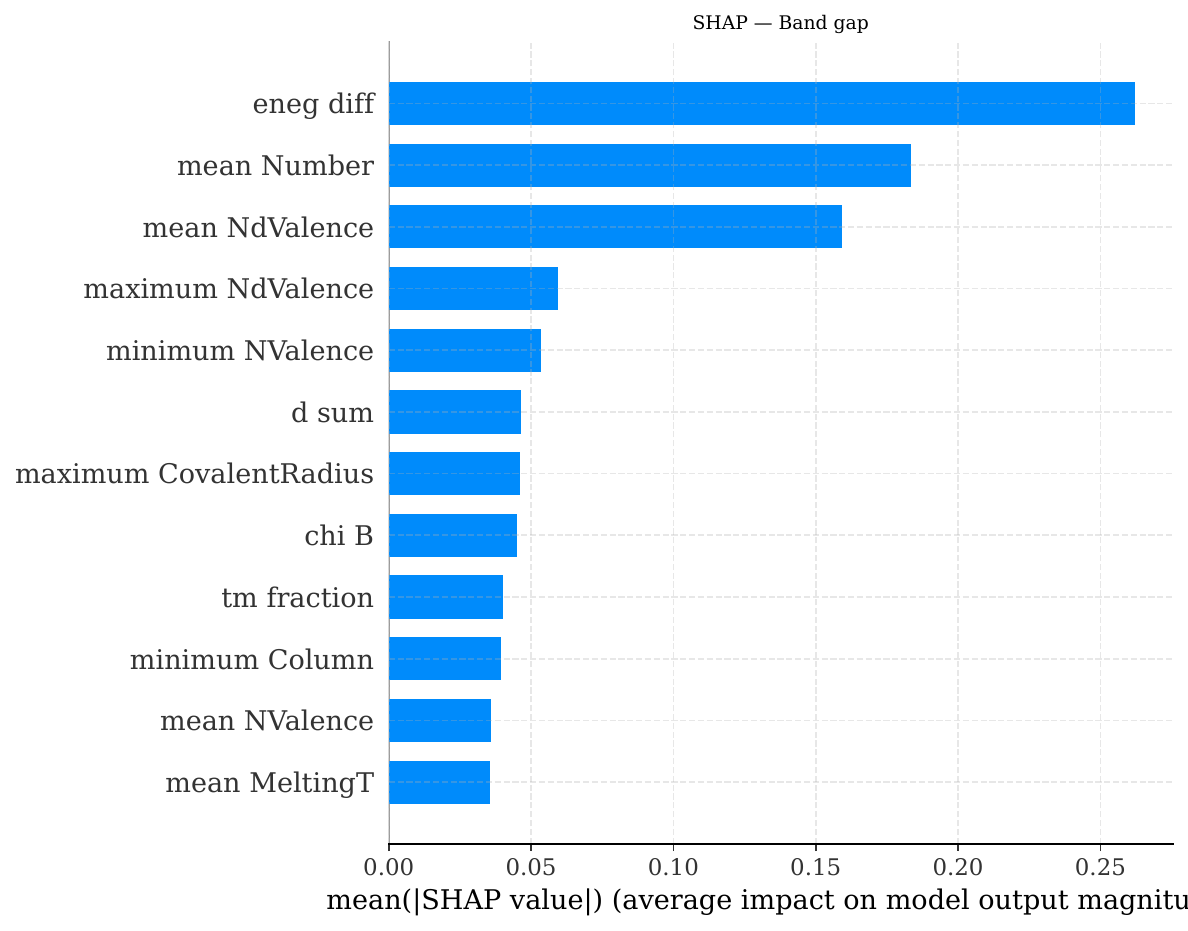}}\hfill
\subfloat[]{\includegraphics[width=0.48\textwidth]{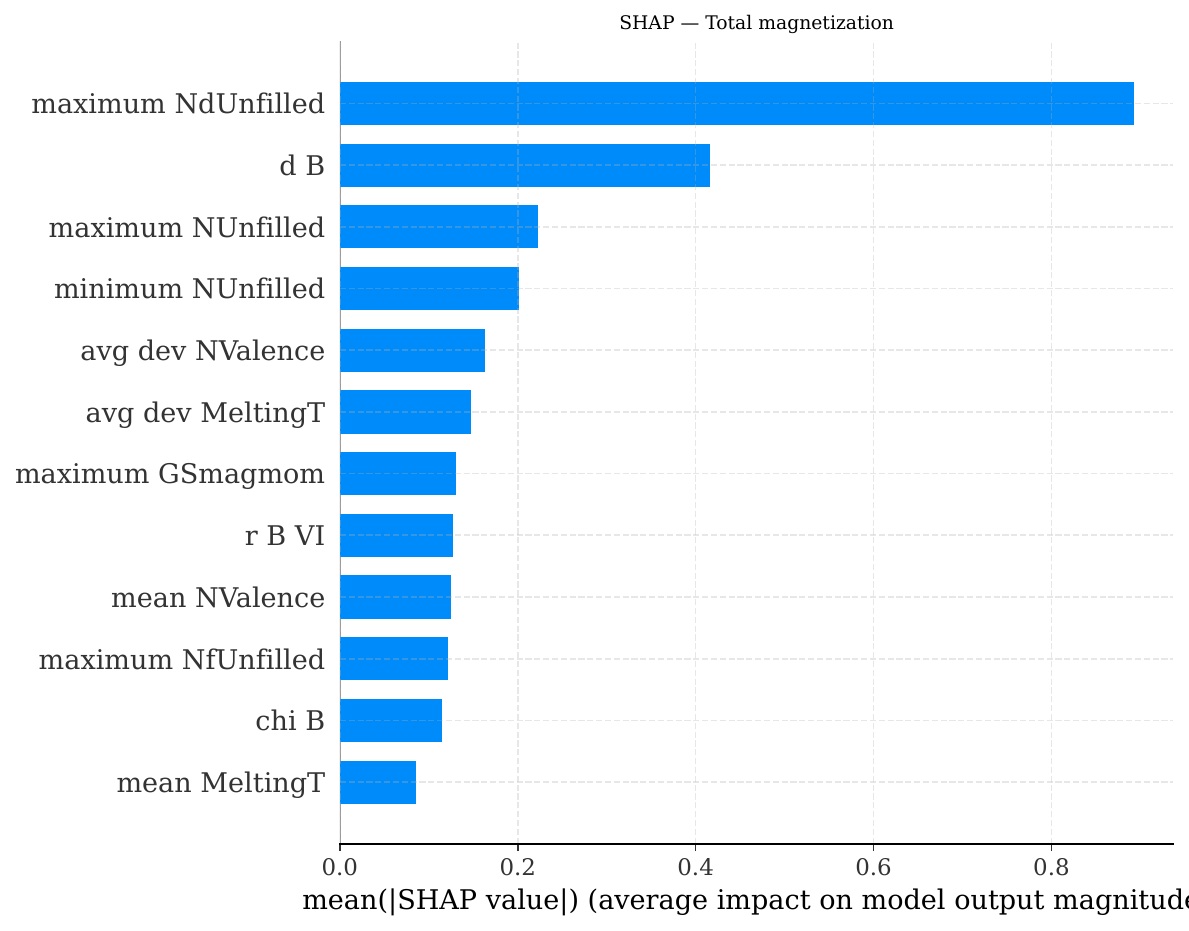}}
\caption{Mean-absolute-SHAP rankings, complementary to
Fig.~\ref{fig:beeswarm}, for (a)~formation energy, (b)~energy above hull,
(c)~band gap, (d)~total magnetization. Numerical top lists for (b)-(d) are in
Table~\ref{tab:shap}, the formation-energy list is archived in the result
bundle.}
\label{fig:shapbar}
\end{figure*}

SHAP~\cite{lundberg2017shap} attribution (Figs.~\ref{fig:beeswarm} and
\ref{fig:shapbar}, Table~\ref{tab:shap}) was computed under the grouped-OOF
protocol, and a feature-group ablation (Fig.~\ref{fig:ablation},
Table~\ref{tab:ablation}) quantifies what the bespoke descriptors add.

\paragraph{Formation energy.}
The anion-cation electronegativity contrast $\Delta\chi$ and the octahedral
electronegativity $\chi_B$ lead the ranking [Figs.~\ref{fig:beeswarm}(a) and
\ref{fig:shapbar}(a)], ahead of covalent-radius and $d$-valence statistics;
the full numerical list is archived in the result bundle. That
electronegativity descriptors carry most of the formation-energy signal is the
ionicity-driven stabilization picture, and the symbolic search of
Sec.~\ref{sec:symbolic} recovers it independently: its closed-form expression
is built from the same two quantities.

\begin{table}[t]
\centering
\caption{Feature-group ablation (grouped-OOF protocol): cumulative addition of
the 19 bespoke structural/electronic descriptors and of the 3-descriptor
occupancy block (Jahn-Teller flag, occupancy proxy, coordination-number gap)
to the Magpie baseline.}
\label{tab:ablation}
\footnotesize
\begin{tabular}{@{}l l c c c c@{}}
\toprule
Target & Feature set & $n_{\mathrm{feat}}$ & MAE & RMSE & $R^2$ \\
\midrule
\multirow{3}{*}{Formation energy}
 & Magpie only  & 120 & 0.1457 & 0.2738 & 0.8954 \\
 & + spinel     & 139 & 0.1253 & 0.2585 & 0.9068 \\
 & + occupancy  & 142 & 0.1272 & 0.2607 & 0.9052 \\
\midrule
\multirow{3}{*}{Energy above hull}
 & Magpie only  & 120 & 0.0497 & 0.1075 & 0.2889 \\
 & + spinel     & 139 & 0.0470 & 0.1054 & 0.3166 \\
 & + occupancy  & 142 & 0.0469 & 0.1052 & 0.3185 \\
\midrule
\multirow{3}{*}{Band gap}
 & Magpie only  & 120 & 0.5095 & 0.7666 & 0.5841 \\
 & + spinel     & 139 & 0.4881 & 0.7346 & 0.6181 \\
 & + occupancy  & 142 & 0.4939 & 0.7373 & 0.6153 \\
\bottomrule
\end{tabular}
\end{table}

\paragraph{Band gap.}
The anion-cation electronegativity contrast dominates (mean $|$SHAP$|$
0.262), followed by the mean atomic number (0.183) and the mean $d$-valence
count (0.159), $\chi_B$ (0.045) and the transition-metal fraction (0.040) also
rank in the top group. This ordering fits the expected two-component picture
of the gap in these compounds: bonding ionicity sets the charge-transfer
component, and $d$-orbital localization the Mott-Hubbard component. Given the
weak holdout skill for this target, we read the attribution as a description
of what the model uses, not as evidence that the model is quantitatively
right.

\paragraph{Total magnetization.}
The maximum unfilled-$d$ count dominates (0.892), followed by the octahedral
$d$-count $d_B$ (0.416) and the maximum and minimum unfilled-orbital counts
(0.223 and 0.201), $r_B^{VI}$ (0.127) and $\chi_B$ (0.115) follow. The
beeswarm [Fig.~\ref{fig:beeswarm}(d)] shows a monotonic, sign-consistent
dependence of the moment on unfilled-$d$ occupancy. This is the localized
Hund's-rule moment picture at the octahedral site, recovered by the model from
data.

\paragraph{Energy above hull.}
Here the leading contributors are local-coordination-geometry descriptors: the
mean pentagonal-pyramidal $\mathrm{CN}_6$ motif (0.00376), $\chi_B$ (0.00305),
the maximum ground-state volume per atom (0.00302), and the mean see-saw
$\mathrm{CN}_4$ motif (0.00286). That coordination motifs rather than
composition carry the hull signal is consistent with stability depending on
distortions and site preferences that composition alone does not encode.

\paragraph{Ablation.}
Adding the 19 bespoke descriptors to the 120 Magpie features improves the
grouped-OOF formation-energy MAE from 0.1457 to \SI{0.1253}{eV/atom}, the hull
MAE from 0.0497 to \SI{0.0470}{eV/atom}, and the gap MAE from 0.5095 to
\SI{0.4881}{eV}, the smallest relative gain. Adding the three-descriptor
occupancy block (Jahn-Teller flag, occupancy proxy, coordination-number gap)
changes little and slightly worsens the formation energy and the gap
(Table~\ref{tab:ablation}). So the structural/electronic block carries real,
non-redundant signal, while the occupancy block, as currently constructed,
does not, a continuous inversion descriptor is an obvious next step.

\paragraph{Permutation null.}
Label permutation collapses the mean OOF $R^2$ to $-0.207\pm0.102$ (formation
energy), $-0.147\pm0.023$ (hull), $-0.252\pm0.140$ (gap), and $-0.234\pm0.097$
(magnetization), confirming that the pipeline cannot manufacture skill from
scrambled labels.

\subsection{Transferability under held-out chemistries}
\label{sec:transfer}

\begin{figure}[t]
\centering
\includegraphics[width=0.92\linewidth]{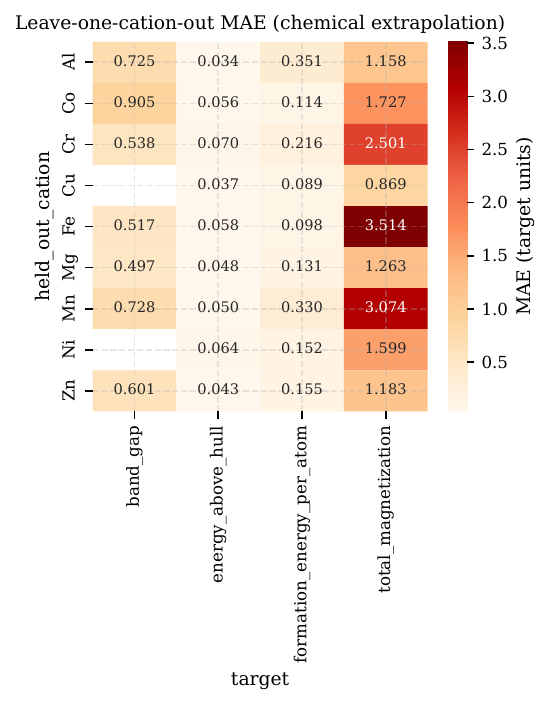}
\caption{Leave-one-cation-out MAE (chemical extrapolation) for each target.
Magnetization extrapolation degrades sharply for the most magnetically active
cations (Fe, Cr, Mn), formation energy and hull distance are more stable. Note
the metric: this map reports MAE, whereas the text and
Table~\ref{tab:robustness} quote $R^2$ ranges, which normalize by the variance
of each withheld subset and can order cations differently
(Sec.~\ref{sec:transfer}). Blank cells mark cations with no held-out band-gap
entries.}
\label{fig:loco}
\end{figure}

LOAO and LOCO results (Fig.~\ref{fig:loco}, ranges in
Table~\ref{tab:robustness}) probe chemical extrapolation, with the
hyperparameter-reuse caveat of Sec.~\ref{sec:robust}. Withholding the small
nitride class gives strongly negative $R^2$ for every target (formation energy
$-1.560$, hull $-0.162$, gap $-0.229$, magnetization $-1.304$). Withholding
oxygen, the majority anion, also fails for the energies (formation energy
$-1.474$, hull $-0.140$): the wide oxide formation-energy range cannot be
recovered from chalcogenides alone. Inside the chalcogenide subspace transfer
is much better, withholding sulfur gives a formation-energy $R^2$ of 0.507 and
withholding selenium gives the best LOAO figures (formation energy 0.870, hull
0.613, magnetization 0.626). Band-gap LOAO is uniformly poor (all
$R^2\le0.018$).

LOCO transfer is uneven by element. Formation-energy extrapolation is
strongest for Cu ($R^2=0.965$) and weakest for Mn (0.614), hull extrapolation
ranges from $-0.381$ (Co) to 0.496 (Cu), magnetization extrapolation reaches
0.803 for Cu but collapses for the most magnetically active cations, Fe
($-0.336$) and Cr ($-1.630$). Figure~\ref{fig:loco} reports the same analysis
as per-cation MAE, because $R^2$ normalizes by the variance of each withheld
subset while MAE does not, the two metrics can order cations
differently-Al, for instance, carries the largest formation-energy MAE in
Fig.~\ref{fig:loco} without being the weakest by $R^2$-so the MAE map and
the $R^2$ ranges of Table~\ref{tab:robustness} should be read together.
Magnetization prediction is acutely sensitive to the local magnetic
environment of the withheld cation, and this bounds how far the surrogate can
be trusted outside its training chemistry.

\subsection{Symbolic regression}
\label{sec:symbolic}

To probe whether a compact analytical relation exists, a gplearn symbolic
search was run over a fixed set of 14 named raw physical descriptors (ionic
radii, radius ratio, tolerance factor, $d$-counts, transition-metal fraction,
oxidation states, electronegativities, and $\Delta\chi$), median-imputed and
otherwise untransformed. gplearn's operators are protected: division guards
small denominators, and $\mathrm{sqrt}$ and $\log$ act on absolute values.
For the formation energy, the search converged to
\begin{equation}
\Eform^{\mathrm{sym}} \simeq \log\chi_B - \Delta\chi
 - \sqrt{\bigl|\,\log\chi_B - \sqrt{|\chi_B - \chi_X|}\,\bigr|},
\label{eq:symbolic}
\end{equation}
where the variable identities follow directly from the fixed column order of
the descriptor list. Validation on one held-out grouped fold gives an MAE of
\SI{0.2758}{eV/atom}. Equation~\eqref{eq:symbolic} should be read as an
empirical surrogate rather than a physical law: it carries no fitted constants
or units, the protected operators matter (for chalcogenide anions
$\chi_X>\chi_B$, so the inner square root acts on $|\chi_B-\chi_X|$), and its
error is two to three times that of the tree ensemble (\SI{0.2758}{eV/atom}
against \SI{0.0865}{eV/atom} on the primary holdout and \SI{0.1206}{eV/atom}
over repeated holdouts). What it does show is that a usable fraction of the
formation-energy signal in this family is carried by electronegativity
descriptors alone, in line with an ionicity-driven stabilization picture and
with the SHAP attribution of the formation-energy champion
(Sec.~\ref{sec:shap}), which ranks the same two electronegativity descriptors
first. For the energy above the hull the search converged to a constant,
\SI{0.016}{eV/atom} (validation MAE \SI{0.0702}{eV/atom}): under this search
budget and parsimony penalty, no low-complexity expression beat a constant,
which matches the SHAP finding that the hull signal lives in nonlinear
coordination-geometry descriptors outside the 14-feature search space.

\subsection{Screening and DFT-validation export}
\label{sec:screening}

\begin{figure*}[htbp]
\centering
\subfloat[]{\includegraphics[width=0.42\textwidth]{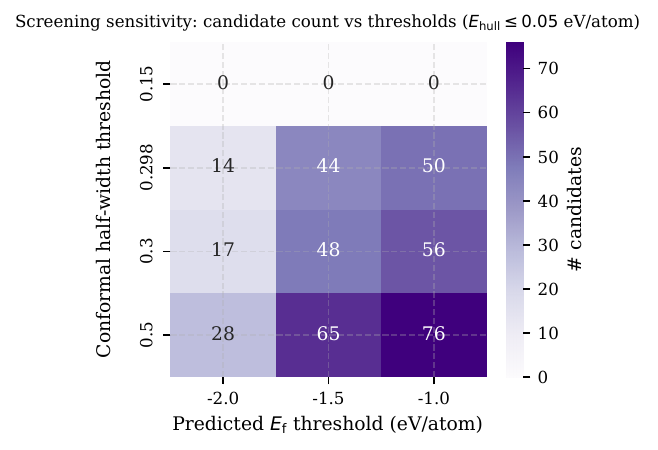}}\hfill
\subfloat[]{\includegraphics[width=0.57\textwidth]{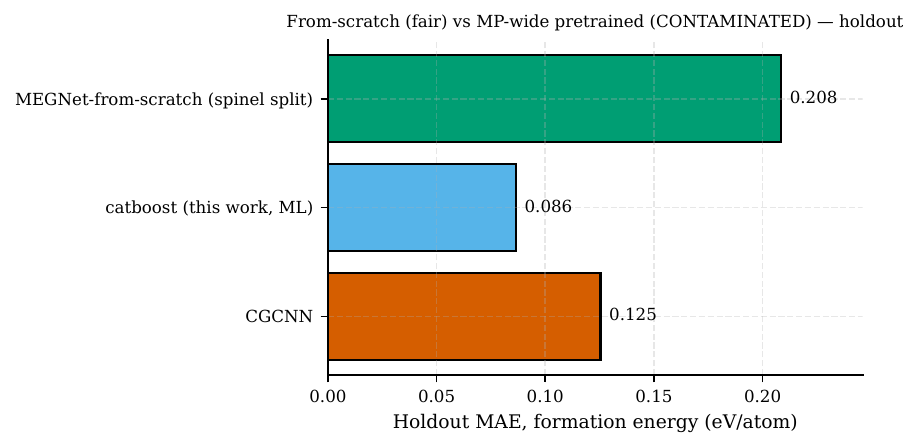}}
\caption{(a)~Screening-funnel sensitivity: candidate count versus the
predicted-$\Eform$ and conformal half-width thresholds at
$\Ehull\le\SI{0.05}{eV/atom}$ (the $p_{\mathrm{metal}}<0.5$ gate is always
applied), nothing survives the tightest width (Table~\ref{tab:screening}).
(b)~Formation-energy holdout MAE of the tabular champion versus the
contaminated pretrained CGCNN bound and the from-scratch MEGNet on the
identical split (values in Table~\ref{tab:pretrained}).}
\label{fig:screening}
\end{figure*}

The stability gate combines four criteria: a predicted-$\Eform$ cutoff, a
predicted-$\Ehull$ cutoff, predicted non-metallicity ($p_{\mathrm{metal}}<0.5$
from the metallicity champion), and an upper bound on the formation-energy
conformal half-width (grouped CV$+$, Sec.~\ref{sec:pipe}). Because those
half-widths are conservative (Sec.~\ref{sec:perf}), the width criterion is
stringent by construction. The candidate-count sensitivity to the three
continuous thresholds is reported in Fig.~\ref{fig:screening}(a) and
Table~\ref{tab:screening}. At the tightest width tested (\SI{0.15}{eV/atom})
nothing passes at any energy cutoff. At the median half-width of the screened
pool (\SI{0.298}{eV/atom}) the count ranges from 14 to 50 depending on the
energy cutoffs, and at the loosest width (\SI{0.5}{eV/atom}) it rises to 76.
Any shortlist is therefore strongly contingent on the uncertainty stringency,
and we report the full sensitivity rather than a single funnel count.

Two exercises were run. The first applies the gate to the curated set itself,
scoring each entry with out-of-fold predictions (training entries) or holdout
predictions (test entries), so that no entry is scored by a model that saw its
label. Twenty entries pass at the median-width stringency, among them
lanthanide chalcogenide spinels such as Ho$_2$MgSe$_4$ (mp-1103791, predicted
$\Ehull=\SI{-0.0043}{eV/atom}$, half-width \SI{0.295}{eV/atom}) and
Tm$_2$CdS$_4$ (mp-4324, predicted \SI{-0.0030}{eV/atom}). Slightly negative
predicted hull distances are artifacts of unconstrained regression and are
read as "on the hull within model error", the elevated half-widths of the
$f$-electron entries correctly flag them as the least certain. Since the true
Materials Project hull distances of these compounds exist, this exercise is a
self-consistency check of the gate, not a discovery claim, and the predicted
values can be compared entry by entry with the database in the result bundle.

The second exercise enumerates 141 charge-neutral $A^{2+}B^{3+}_2X^{2-}_4$
compositions absent from the Materials Project, with
$A\in\{$Ca, Cd, Co, Cu, Fe, Mg, Mn, Ni, Zn$\}$,
$B\in\{$Al, Co, Cr, Fe, Ga, In, Mn, Sc, Ti, V$\}$, and $X\in\{$O, S, Se$\}$.
No relaxed structures exist for these, so they were scored by a
composition-only model: the champion architecture and hyperparameters
retrained on the training partition using Magpie features alone. Three
consequences must be stated plainly. The conformal guarantees of the
exchangeable regime do not extend to this extrapolation. The metallicity
filter of the gate was not applied to this list, $p_{\mathrm{metal}}$ is
reported for information only, and most of the lowest-$\Ehull$ compositions
are in fact predicted metallic (Table~\ref{tab:candidates}). And the ten
lowest predicted hull distances span only \SI{0.02}{eV/atom}, well inside the
model MAE of \SI{0.043}{eV/atom}, so Table~\ref{tab:candidates} is an
unordered shortlist, not a ranking. Absence from the Materials Project also
does not imply novelty: several enumerated compositions are known
experimentally in other structure types (NiCr$_2$Se$_4$, for instance, adopts
the monoclinic Cr$_3$S$_4$ type), and a literature and ICSD check is part of
the intended DFT triage, since the composition-only model cannot see competing
polymorphs.

\begin{table}[th]
\centering
\caption{Screening-funnel sensitivity: number of curated entries passing the
combined gate (predicted $\Eform$, predicted $\Ehull$, predicted
non-metallicity $p_{\mathrm{metal}}<0.5$, and formation-energy conformal
half-width) as a function of the three continuous thresholds. The
0.298~eV/atom row is the median half-width of the screened pool.
Representative rows, full grid in Fig.~\ref{fig:screening}(a).}
\label{tab:screening}
\footnotesize
\begin{tabular}{@{}c c c r@{}}
\toprule
Half-width thr.\ (\si{eV/atom}) & $\Eform$ thr.\ (\si{eV/atom}) & $\Ehull$ thr.\ (\si{eV/atom}) & $n_{\mathrm{cand}}$ \\
\midrule
0.150 & $-1.0$ & 0.05 & 0 \\
0.150 & $-2.0$ & 0.10 & 0 \\
0.298 & $-1.0$ & 0.05 & 50 \\
0.298 & $-1.0$ & 0.10 & 68 \\
0.298 & $-2.0$ & 0.05 & 14 \\
0.298 & $-2.0$ & 0.10 & 23 \\
0.300 & $-1.0$ & 0.05 & 56 \\
0.300 & $-1.0$ & 0.10 & 75 \\
0.500 & $-1.0$ & 0.05 & 76 \\
0.500 & $-1.0$ & 0.10 & 122 \\
\bottomrule
\end{tabular}
\end{table}

\begin{table}[th]
\centering
\caption{Ten generated $AB_2X_4$ compositions (absent from the Materials
Project) with the lowest composition-only predicted hull distance. The spread
of $\Ehull^{\mathrm{pred}}$ across the table (\SI{0.02}{eV/atom}) is well
inside the model MAE (\SI{0.043}{eV/atom}), so the list is an unordered
shortlist, not a ranking. No metallicity filter was applied;
$p_{\mathrm{metal}}$ is informational. Predictions carry no conformal
guarantee in this extrapolation regime, and absence from the database does not
imply novelty (Sec.~\ref{sec:screening}). DFT-ready prototypes
(POSCAR/INCAR/KPOINTS) were emitted for each entry.}
\label{tab:candidates}
\footnotesize
\begin{tabular}{@{}l l c c c@{}}
\toprule
 & Formula & $\Eform^{\mathrm{pred}}$ (\si{eV/atom}) & $\Ehull^{\mathrm{pred}}$ (\si{eV/atom}) & $p_{\mathrm{metal}}$ \\
\midrule
 & MnCr$_2$Se$_4$   & $-0.806$ & $-0.0016$ & 0.971 \\
 & Co$_2$NiSe$_4$   & $-0.369$ & $0.0021$  & 0.981 \\
 & Ti$_2$CuSe$_4$   & $-1.080$ & $0.0082$  & 0.979 \\
 & Sc$_2$CuS$_4$    & $-1.586$ & $0.0138$  & 0.936 \\
 & Sc$_2$NiSe$_4$   & $-1.544$ & $0.0146$  & 0.915 \\
 & Mn$_2$CuSe$_4$   & $-0.724$ & $0.0156$  & 0.989 \\
 & Cr$_2$NiSe$_4$   & $-0.663$ & $0.0164$  & 0.981 \\
 & Ti$_2$CdSe$_4$   & $-1.218$ & $0.0165$  & 0.743 \\
 & Sc$_2$CuSe$_4$   & $-1.483$ & $0.0165$  & 0.913 \\
 & MgTi$_2$Se$_4$   & $-1.460$ & $0.0182$  & 0.205 \\
\bottomrule
\end{tabular}
\end{table}

For the ten shortlisted compositions, prototype structures were built by
cation substitution onto an isostructural, same-anion template from the
training set, and POSCAR files with template INCAR/KPOINTS inputs were emitted
for DFT$+U$ relaxation: \SI{520}{eV} plane-wave cutoff, $10^{-6}$~eV energy
convergence and \SI{0.01}{eV/\angstrom} force convergence
(\texttt{EDIFFG}$\,=-0.01$), full ionic and cell relaxation, spin
polarization, Dudarev DFT$+U$ with per-species $U$ left for manual
specification, and a $6\times6\times6$ $\Gamma$-centered mesh.

Finally, Fig.~\ref{fig:screening}(b) places the tabular champion next to the
graph-network references. The champion (\SI{0.0865}{eV/atom} on the primary
holdout) is more accurate than the contaminated pretrained CGCNN bound
(\SI{0.1254}{eV/atom}), and the from-scratch MEGNet on the identical split
reaches \SI{0.2085}{eV/atom} ($R^2=0.888$) for the formation energy and
\SI{0.7520}{eV} ($R^2=-0.190$) for the gap on the non-metal subset, no better
than the dummy baseline there (\SI{0.7404}{eV}). With 272 training compounds,
a message-passing network cannot amortize its parameter count, while the
descriptors inject the coordination chemistry it would otherwise have to
learn. The scope of this conclusion is limited by the single untuned MEGNet
configuration, as discussed below.


\section{Limitations}
\label{sec:limitations}

Several limitations bound the interpretation. (i)~All labels derive from the
Materials Project's mixed GGA/GGA$+U$ scheme with compatibility
corrections~\cite{wang2021corrections}, the gaps in particular are semi-local
and severely underestimated for Mott and charge-transfer oxides, and the
metallicity label inherits that bias, only partially mitigated by the
threshold study of Sec.~\ref{sec:perf}. (ii)~The magnetization label is
whatever configuration the underlying calculation converged to from a
ferromagnetic start, ferrimagnetic and non-collinear ground states are not
guaranteed. (iii)~The set is oxide-heavy, and both the nitride minority and
the heterogeneous oxide majority extrapolate poorly under LOAO, broader anion
sampling would help. (iv)~The energy targets are near their data-limited
plateau while the gap is not, so the gap result is a power-limited negative,
and the 19-member non-metal holdout gives wide coverage intervals.
(v)~Several screened entries are $f$-electron lanthanide spinels, for which
both the descriptor set and the DFT$+U$ ground truth are least reliable.
(vi)~The generative screen is composition-only, outside the exchangeability
assumption of the conformal construction, and unfiltered for electronic
character. (vii)~Hull distances are 0~K electronic-energy quantities;
vibrational, configurational, and finite-temperature contributions are
neglected, and DFT$+U$ results depend on the chosen $U$. (viii)~The
from-scratch MEGNet used one fixed configuration, one seed, and a fixed
80-epoch budget with no early stopping and no hyperparameter search, whereas
the tabular models were tuned, the \SI{0.2085}{eV/atom} figure is a
representative small-data GNN reference, not an optimized one, and a tuned or
transfer-learned network would likely narrow, though at this sample size
probably not close, the gap. (ix)~The out-of-fold statistics are non-nested
with respect to tuning, and the repeated-holdout and leave-one-chemistry-out
analyses reuse hyperparameters tuned on the primary training partition, both
are mildly optimistic, which is why the grouped holdout remains the primary
evidence. (x)~The grouped conformal construction over-covers for the formation
energy (empirical coverage 1.00 at nominal 0.90, Sec.~\ref{sec:perf}), and the
width-gated screen inherits that conservatism, better-calibrated intervals
(Mondrian or residual-normalized conformal, for instance) would loosen the
gate without changing the underlying models. All conclusions rest on
computational surrogates trained on computational ground truth, synthesis,
magnetometry, and diffraction of any shortlisted candidate are required before
a prediction becomes a design recommendation.

\section{Conclusion}

We have applied a coordination-resolved, group-aware surrogate framework to
320 cubic spinel nitrides, oxides, sulfides, and selenides from the Materials
Project, with CrystalNN-based site assignment in place of element-identity
rules and with every predictive claim conditioned on beating trivial and
linear baselines under paired bootstrap testing and surviving a
label-permutation null. The outcome is differentiated, deliberately so.
Formation energy, hull distance, and magnetization are predicted with
statistically significant accuracy-repeated-holdout MAEs of
$0.121\pm0.030$~eV/atom, $0.048\pm0.013$~eV/atom, and $1.27\pm0.19$~\muBfu{},
with the stated caveat that the repeats reuse the tuned
hyperparameters-and metallicity is classified with an AUC of 0.93. Band-gap
regression does not beat the trivial baseline on the small non-metal holdout,
a power-limited negative result on semi-local labels rather than evidence of
a ceiling. On the identical grouped split the tabular champion outperforms an
untuned, single-configuration MEGNet trained from scratch, and it also beats
the contaminated pretrained CGCNN bound, both comparisons support
descriptor-based models for narrow structure families at this data scale,
while leaving open how much a tuned or transfer-learned network would recover.
The attribution is physically sensible where the models are skillful:
unfilled-$d$ occupancy at the octahedral site carries the moment, and
electronegativity descriptors carry the formation energy-a picture the
symbolic-regression expression recovers independently-as well as what
little gap signal exists. The ablation shows genuine, non-redundant signal in
the bespoke structural block, while the current binary occupancy proxy adds
nothing, a continuous inversion descriptor is the natural next step.
Transferability is uneven: chalcogenide-to-chalcogenide extrapolation works,
nitrides and the most magnetically active cations do not. The conformal
intervals are honest but conservative-the formation-energy intervals
over-cover at roughly seven times the point error, and it is these widths
that make the screening gate stringent-so the hull model is deployed only
as a coarse, sensitivity-reported filter, and the composition-only generative
shortlist is presented unordered, with DFT-ready inputs, as a starting point
for first-principles validation rather than as discovery.

\section*{Data and Code Availability}

The data and source code supporting the findings of this study are
available from the corresponding author upon reasonable request. The
complete codebase, datasets, analysis scripts, and documentation will be
made publicly available in a GitHub repository following publication of
this work.

\section*{Author Contributions}

\textbf{K. KHALLOUQ:} Conceptualization, Methodology, Investigation, Data Curation,
Formal Analysis, Visualization, Writing \& Editing.

\textbf{A. EL MAAZOUZI:} Conceptualization, Methodology, Software, Formal Analysis, Visualization, 
Writing, Review \& Editing.

\textbf{R. MASROUR:} Supervision, Validation, Investigation,
Writing, Review \& Editing.



\bibliographystyle{apsrev4-2}
\bibliography{spinel_V6_3}

@article{Maazouzi2019,
  title = {Computational study of inverse ferrite spinels},
  volume = {28},
  ISSN = {1674-1056},
  url = {http://dx.doi.org/10.1088/1674-1056/28/5/057504},
  DOI = {10.1088/1674-1056/28/5/057504},
  number = {5},
  journal = {Chinese Physics B},
  publisher = {IOP Publishing},
  author = {Maazouzi,  A EL and Masrour,  R and Jabar,  A and Hamedoun,  M},
  year = {2019},
  month = May,
  pages = {057504}
}

@article{Khallouq2023,
  title = {Magnetic properties and magnetocaloric effect on ACr2Se4 (A = Hg and Cd): a Monte Carlo study},
  volume = {97},
  ISSN = {0974-9845},
  url = {http://dx.doi.org/10.1007/s12648-023-02693-0},
  DOI = {10.1007/s12648-023-02693-0},
  number = {12},
  journal = {Indian Journal of Physics},
  publisher = {Springer Science and Business Media LLC},
  author = {Khallouq,  K. and Masrour,  R. and Maazouzi,  A. El},
  year = {2023},
  month = Apr,
  pages = {3515–3522}
}

@article{anderson1950,
  author  = {Anderson, P. W.},
  title   = {Antiferromagnetism. {T}heory of Superexchange Interaction},
  journal = {Physical Review},
  year    = {1950},
  volume  = {79},
  pages   = {350--356},
  doi     = {10.1103/PhysRev.79.350}
}

@article{goodenough1955,
  author  = {Goodenough, J. B.},
  title   = {Theory of the Role of Covalence in the Perovskite-Type Manganites [{L}a, {M}(II)]{M}n{O}3},
  journal = {Physical Review},
  year    = {1955},
  volume  = {100},
  pages   = {564--573},
  doi     = {10.1103/PhysRev.100.564}
}

@article{kanamori1959,
  author  = {Kanamori, Junjiro},
  title   = {Superexchange interaction and symmetry properties of electron orbitals},
  journal = {Journal of Physics and Chemistry of Solids},
  year    = {1959},
  volume  = {10},
  number  = {2-3},
  pages   = {87--98},
  doi     = {10.1016/0022-3697(59)90061-7}
}

@article{zener1951,
  author  = {Zener, Clarence},
  title   = {Interaction between the $d$-Shells in the Transition Metals. {II}. {F}erromagnetic Compounds of Manganese with Perovskite Structure},
  journal = {Physical Review},
  year    = {1951},
  volume  = {82},
  pages   = {403--405},
  doi     = {10.1103/PhysRev.82.403}
}

@article{oneill1983,
  author  = {O'Neill, Hugh St. C. and Navrotsky, Alexandra},
  title   = {Simple spinels: crystallographic parameters, cation radii, lattice energies, and cation distribution},
  journal = {American Mineralogist},
  year    = {1983},
  volume  = {68},
  number  = {1-2},
  pages   = {181--194},
}

@article{jain2016review,
  author  = {Jain, Anubhav and Shin, Yongwoo and Persson, Kristin A.},
  title   = {Computational predictions of energy materials using density functional theory},
  journal = {Nature Reviews Materials},
  year    = {2016},
  volume  = {1},
  pages   = {15004},
  doi     = {10.1038/natrevmats.2015.4}
}

@article{kirklin2015,
  author  = {Kirklin, Scott and Saal, James E. and Meredig, Bryce and Thompson, Alex and Doak, Jeff W. and Aykol, Muratahan and R{\"u}hl, Stephan and Wolverton, Chris},
  title   = {The Open Quantum Materials Database ({OQMD}): assessing the accuracy of {DFT} formation energies},
  journal = {npj Computational Materials},
  year    = {2015},
  volume  = {1},
  pages   = {15010},
  doi     = {10.1038/npjcompumats.2015.10}
}

@article{perdew1985,
  author  = {Perdew, John P.},
  title   = {Density functional theory and the band gap problem},
  journal = {International Journal of Quantum Chemistry},
  year    = {1985},
  volume  = {28},
  pages   = {497--523},
  doi     = {10.1002/qua.560280846}
}

@article{wang2021corrections,
  author  = {Wang, Amanda and Kingsbury, Ryan and McDermott, Matthew and Horton, Matthew and Jain, Anubhav and Ong, Shyue Ping and Dwaraknath, Shyam and Persson, Kristin A.},
  title   = {A framework for quantifying uncertainty in {DFT} energy corrections},
  journal = {Scientific Reports},
  year    = {2021},
  volume  = {11},
  pages   = {15496},
  doi     = {10.1038/s41598-021-94550-5}
}

@article{dunn2020matbench,
  author  = {Dunn, Alexander and Wang, Qi and Ganose, Alex and Dopp, Daniel and Jain, Anubhav},
  title   = {Benchmarking materials property prediction methods: the {M}atbench test set and {A}utomatminer reference algorithm},
  journal = {npj Computational Materials},
  year    = {2020},
  volume  = {6},
  pages   = {138},
  doi     = {10.1038/s41524-020-00406-3}
}

@article{bartel2020,
  author  = {Bartel, Christopher J. and Trewartha, Amalie and Wang, Qi and Dunn, Alexander and Jain, Anubhav and Ceder, Gerbrand},
  title   = {A critical examination of compound stability predictions from machine-learned formation energies},
  journal = {npj Computational Materials},
  year    = {2020},
  volume  = {6},
  pages   = {97},
  doi     = {10.1038/s41524-020-00362-y}
}

@article{jain2013mp,
  author  = {Jain, Anubhav and Ong, Shyue Ping and Hautier, Geoffroy and Chen, Wei and Richards, William Davidson and Dacek, Stephen and Cholia, Shreyas and Gunter, Dan and Skinner, David and Ceder, Gerbrand and Persson, Kristin A.},
  title   = {Commentary: {T}he {M}aterials {P}roject: {A} materials genome approach to accelerating materials innovation},
  journal = {APL Materials},
  year    = {2013},
  volume  = {1},
  pages   = {011002},
  doi     = {10.1063/1.4812323}
}

@article{zimmermann2020,
  author  = {Zimmermann, Nils E. R. and Jain, Anubhav},
  title   = {Local structure order parameters and site fingerprints for quantification of coordination environment and crystal structure similarity},
  journal = {RSC Advances},
  year    = {2020},
  volume  = {10},
  pages   = {6063--6081},
  doi     = {10.1039/C9RA07755C}
}

@article{ward2016magpie,
  author  = {Ward, Logan and Agrawal, Ankit and Choudhary, Alok and Wolverton, Christopher},
  title   = {A general-purpose machine learning framework for predicting properties of inorganic materials},
  journal = {npj Computational Materials},
  year    = {2016},
  volume  = {2},
  pages   = {16028},
  doi     = {10.1038/npjcompumats.2016.28}
}

@article{xie2018cgcnn,
  author  = {Xie, Tian and Grossman, Jeffrey C.},
  title   = {Crystal Graph Convolutional Neural Networks for an Accurate and Interpretable Prediction of Material Properties},
  journal = {Physical Review Letters},
  year    = {2018},
  volume  = {120},
  pages   = {145301},
  doi     = {10.1103/PhysRevLett.120.145301}
}

@article{chen2019megnet,
  author  = {Chen, Chi and Ye, Weike and Zuo, Yunxing and Zheng, Chen and Ong, Shyue Ping},
  title   = {Graph Networks as a Universal Machine Learning Framework for Molecules and Crystals},
  journal = {Chemistry of Materials},
  year    = {2019},
  volume  = {31},
  number  = {9},
  pages   = {3564--3572},
  doi     = {10.1021/acs.chemmater.9b01294}
}

@inproceedings{lundberg2017shap,
  author    = {Lundberg, Scott M. and Lee, Su-In},
  title     = {A Unified Approach to Interpreting Model Predictions},
  booktitle = {Advances in Neural Information Processing Systems 30 (NeurIPS 2017)},
  year      = {2017},
  eprint    = {1705.07874},
  archivePrefix = {arXiv},
  doi     = {10.48550/arXiv.1705.07874}
}

@article{pedregosa2011sklearn,
  author  = {Pedregosa, Fabian and Varoquaux, Ga{\"e}l and Gramfort, Alexandre and Michel, Vincent and Thirion, Bertrand and Grisel, Olivier and Blondel, Mathieu and Prettenhofer, Peter and Weiss, Ron and Dubourg, Vincent and Vanderplas, Jake and Passos, Alexandre and Cournapeau, David and Brucher, Matthieu and Perrot, Matthieu and Duchesnay, {\'E}douard},
  title   = {Scikit-learn: Machine Learning in {P}ython},
  journal = {Journal of Machine Learning Research},
  year    = {2011},
  volume  = {12},
  pages   = {2825--2830},
  doi     = {10.48550/arXiv.1201.0490}  
}

@inproceedings{chen2016xgboost,
  author    = {Chen, Tianqi and Guestrin, Carlos},
  title     = {{XGB}oost: A Scalable Tree Boosting System},
  booktitle = {Proceedings of the 22nd ACM SIGKDD International Conference on Knowledge Discovery and Data Mining},
  year      = {2016},
  pages     = {785--794},
  doi       = {10.1145/2939672.2939785}
}

@inproceedings{ke2017lightgbm,
  author    = {Ke, Guolin and Meng, Qi and Finley, Thomas and Wang, Taifeng and Chen, Wei and Ma, Weidong and Ye, Qiwei and Liu, Tie-Yan},
  title     = {{L}ight{GBM}: A Highly Efficient Gradient Boosting Decision Tree},
  booktitle = {Advances in Neural Information Processing Systems 30 (NeurIPS 2017)},
  year      = {2017},
  pages     = {3146--3154}
}

@misc{dorogush2018catboost,
  author = {Dorogush, Anna Veronika and Ershov, Vasily and Gulin, Andrey},
  title  = {{C}at{B}oost: gradient boosting with categorical features support},
  year   = {2018},
  eprint = {1810.11363},
  archivePrefix = {arXiv},
  doi    = {10.48550/arXiv.1810.11363}
}

@article{PhysRev.98.391,
  title = {Theory of Ionic Ordering, Crystal Distortion, and Magnetic Exchange Due to Covalent Forces in Spinels},
  author = {Goodenough, J. B. and Loeb, A. L.},
  journal = {Phys. Rev.},
  volume = {98},
  issue = {2},
  pages = {391--408},
  numpages = {0},
  year = {1955},
  month = {Apr},
  publisher = {American Physical Society},
  doi = {10.1103/PhysRev.98.391},
  url = {https://link.aps.org/doi/10.1103/PhysRev.98.391}
}

\end{document}